\documentclass[prd,aps,10pt,superscriptaddress,twocolumn,amsmath,amssymb,nofootinbib,longbibliography, floatfix]{revtex4-1}

\usepackage{ragged2e}

\usepackage[english]{babel}

\usepackage{natbib}
\usepackage[colorlinks]{hyperref}
\hypersetup{linkcolor=blue,citecolor=blue,filecolor=blue,urlcolor=blue}

\usepackage{amssymb,amsmath,multirow}
\usepackage{bm}
\usepackage{dsfont}
\usepackage{orcidlink}
\usepackage{color}
\usepackage{array}

\usepackage[utf8]{inputenc}
\usepackage{times}
\usepackage{setspace}
\usepackage{float}
\usepackage{caption}
\usepackage{booktabs}
\usepackage{multirow}
\usepackage{siunitx}

\usepackage[normalem]{ulem}
\usepackage{comment}

\usepackage{subfig,hyperref}

\usepackage{graphicx}
\usepackage{dcolumn}
\usepackage{bm}
\usepackage{wasysym}
\usepackage[
   starfontserif 
    ]{starfont}
\usepackage{amsmath}

\usepackage{calligra}

\DeclareMathAlphabet{\mathcalligra}{T1}{calligra}{m}{n}
\DeclareFontShape{T1}{calligra}{m}{n}{<->s*[2.2]callig15}{}
\newcommand{\scriptr}{\mathcalligra{r}\,}

\DeclareSymbolFont{starfontsym}{OT1}{sts}{m}{n}
\DeclareMathSymbol{\mathTerra}{\mathord}{starfontsym}{76}

\newcommand{\beq}{\begin{equation}}
\newcommand{\beqn}{\begin{align}}
\newcommand{\eeq}{\end{equation}}
\newcommand{\eeqn}{\end{align}}

\begin{document}

\title{Electromagnetic Signatures From Primordial Black Holes in the Solar System 
}

\author{Alexandra P. Klipfel \orcidlink{0000-0002-1907-7468}}
 \email{aklipfel@mit.edu}
\affiliation{%
Department of Physics, Massachusetts Institute of Technology, Cambridge, MA 02139, USA}
 
\author{David I.~Kaiser \orcidlink{0000-0002-5054-6744}}
 \email{dikaiser@mit.edu}

\affiliation{%
Department of Physics, Massachusetts Institute of Technology, Cambridge, MA 02139, USA}

\date{\today}

\begin{abstract}
Primordial black holes (PBHs) in the asteroid-mass range, with typical masses $10^{17}\,{\rm g}\lesssim M \lesssim 10^{23}\,{\rm g}$, have drawn significant recent attention as a viable dark matter candidate. The peak frequencies of photons emitted via Hawking radiation from asteroid-mass PBHs range from infrared to $\gamma$-ray bands. We calculate expected local transit rates for extended PBH mass distributions which could comprise all the dark matter.  We evaluate prospects for detecting Hawking-radiated photons from local PBH transits through the inner Solar System and from PBH explosions in the far outer edges of the Solar System. We consider several existing and proposed ground-based and space-based instruments sensitive to photons from the radio band to ultrahigh energy $\gamma$-rays. We find that proposed instruments, such as the All-sky Medium Energy Gamma-ray Observatory eXplorer (AMEGO-X) satellite, could reliably detect PBH transits within ${\cal O} (0.1 \, {\rm AU})$ of the Earth, while the High Altitude Water Cherenkov (HAWC) observatory and Large High-Altitude Air Shower Observatory (LHAASO) are both sensitive to PBH explosions out to $\mathcal{O}(0.1 \, {\rm pc})$ and $\mathcal{O}(0.5 \, {\rm pc})$ respectively. We conclude by specifically considering potential companion electromagnetic signatures in the case of a PBH explosion about $10^3\,{\rm AU}$ from Earth, which has been suggested as a potential source for the $\sim 220 \, {\rm PeV}$ ultrahigh-energy KM3-230213A neutrino event observed by the KM3NeT collaboration in 2023. Whereas we find that the recent KM3NeT event would not have yielded detectable electromagnetic signals---due to its location on the sky, proposed distance from Earth, and the offline status of the HAWC observatory at that time---we demonstrate that future PBH explosions at comparable distances could yield measurable electromagnetic signals at Earth, depending on alignment of the PBH burst with detector fields of view.
\end{abstract}

\maketitle

\section{Introduction}

We consider prospects for detecting electromagnetic (EM) signals of Hawking emission \cite{Hawking:1974rv,Hawking:1975vcx,Page:1976df,Page:1976ki,Page:1977um,macgibbonQuarkGluonjetEmission1990,macgibbonQuarkGluonjetEmission1991} from primordial black holes (PBHs) within the Solar System. The Hawking temperature of a black hole varies inversely with its mass \cite{Hawking:1974rv,Hawking:1975vcx}. Whereas a black hole will emit particles at a relatively steady rate over most of its lifetime, it will eventually reach a phase of rapid, high-energy emission during which the increasing mass-loss rate rapidly increases the temperature in a runaway process that Hawking dubbed an ``explosion'' \cite{Hawking:1974rv}. Hence we consider EM signatures from PBHs in two distinct regimes: (1) steady-state Hawking radiation of photons from PBHs with masses $M \gtrsim 10^{17} \, {\rm g}$ transiting through the Solar System, and (2) time-varying bursts of energetic photons from exploding PBHs whose masses have evolved to $M \lesssim 10^{11} \, {\rm g}$.

We define the Solar System to include the Kuiper Belt and Oort Cloud, which extend to $\mathcal{O}(10^3\,{\rm AU})$ and $\mathcal{O}(10^5\,{\rm AU})$ respectively. We assume that the PBHs are drawn from a galactic PBH population that obeys an extended mass distribution function. We focus particularly on distributions with typical mass $\bar{M}$ peaked in the so-called \emph{asteroid-mass window}, which corresponds to $10^{17}\,{\rm g}\lesssim \bar{M}\lesssim 10^{23}\,{\rm g}$, within which PBHs could comprise all the dark matter. See Refs.~\cite{Khlopov:2008qy,carrConstraintsPrimordialBlack2021,Carr:2020xqk,greenPrimordialBlackHoles2021,Escriva:2022duf,gortonHowOpenAsteroidmass2024,carrObservationalEvidencePrimordial2024,Khan:2025kag} for reviews of the current constraints on the PBH dark matter fraction. 

Detecting photon emission from a PBH transiting through the inner Solar System could provide a ``multimessenger'' counterpart to complement other local PBH observables, such as gravitational perturbations~\cite{tranCloseEncountersPrimordial2024,Cuadrat-Grzybowski:2024uph,Thoss:2024vae,DeLorenci:2025wbn,Thoss:2025yht} or Hawking-emitted positrons~\cite{Klipfel:2025bvh}. Given the local dark matter mass density near the Sun and an expected relative velocity of $\sim250\,{\rm km/s}$, one can expect PBHs to transit within $5\,{\rm AU}$ of the Earth (out to the orbit of Jupiter) at rates ranging from roughly once per century to ten times per year across much of the asteroid-mass range ~\cite{tranCloseEncountersPrimordial2024,Klipfel:2025bvh}.
Moreover, Ref.~\cite{tranCloseEncountersPrimordial2024} shows that a PBH transit with closest approach to the Sun of $R \lesssim 5$ AU would yield measurable perturbations to the motion of well-tracked objects, such as the planet Mars. Therefore we will focus on EM signatures from PBH transits within $R \lesssim 5$ AU of the Sun.

Meanwhile, in Ref.~\cite{Klipfel:2025jql} we demonstrated that realistic PBH mass distributions predict an $\mathcal{O}(8\%)$ chance for a PBH to explode within ${\cal O}(10^3\,{\rm AU})$ of the Earth every 14 years. (See also Ref.~\cite{Baker:2025zxm}.) Such explosions could account for the detection of rare, ultrahigh-energy particles such as neutrinos~\cite{Boccia:2025hpm,Klipfel:2025jql,Baker:2025zxm,Baker:2025cff,Anchordoqui:2025xug,Airoldi:2025opo,Airoldi:2025bgr,Ambrosone:2026djo,Mukhopadhyay:2026lmz}. Since exploding PBHs should emit all particles that exist \cite{Baker:2021btk,Baker:2022rkn,Perez-Gonzalez:2025try} (at rates that depend on the particles' spins and masses), here we focus on detection prospects for companion EM signals that would accompany such ultrahigh-energy neutrino detections for PBH explosions that occur at distances $\mathcal{O}(10^3\,{\rm AU})$ from the Earth. As a concrete example, we analyze whether the recent event KM3-230213A, with $E_\nu \sim 220 \, {\rm PeV}$ \cite{aielloObservationUltrahighenergyCosmic2025}, would have yielded a detectable EM companion signature, if---as suggested in Ref.~\cite{Klipfel:2025jql}---the KM3NeT event had been sourced by a PBH exploding within the Oort Cloud. We find that whereas that particular event would {\it not} be expected to have yielded a detectable EM signature (given its far distance from Earth, and the fact that one of the major $\gamma$-ray detectors happened to be off-line at the time), future PBH explosions at comparable distances {\it could} yield measurable EM signals, depending on chance alignment of the PBH locations and the orientations of various detectors' fields of view at the appropriate times.

Throughout this analysis, we expand upon previous work studying the detection prospects for PBH EM signatures~\cite{sobrinhoDirectDetectionBlack2014,Arbey:2020urq,Auffinger:2022dic,Auffinger:2022khh,Baker:2025ffi} to include modern instruments such as the High-Altitude Water Cherenkov (HAWC) observatory, Large High Altitude Air Shower (LHAASO) observatory, and prospective All-sky Medium Energy Gamma-Ray Observatory eXplorer (AMEGO-X). Given the broad frequency range of Hawking emission, we also consider possibilities for radio-band detection. We use the latest monochromatic PBH evaporation bounds on $f_{\rm PBH}$, the fraction of dark matter energy density that could consist of PBHs \cite{DelaTorreLuque:2024qms,Balaji:2025afr}, to compute constraints on $f_{\rm PBH}$ for realistic extended mass functions. We compute full secondary Hawking spectra numerically with \texttt{BlackHawk v2.2} \cite{arbeyBlackHawkV20Public2019, arbeyPhysicsStandardModel2021a}, and consider how relative motion between the PBH and the detector affects detection prospects. Recent work investigated the possibility of a separate, novel EM signature from PBHs gravitationally ionizing neutral hydrogen in the interplanetary medium \cite{Klipfel:2026aug}. Given that Hawking emission would dominate any such signatures \cite{Klipfel:2026aug}, we neglect any effects from gravitational ionization when modeling the measured photon spectra from local PBH transits and explosions.

\begin{table*}[t]
\caption{\label{tab:HawkIRUV} \justifying Photon energy bands corresponding to various portions of the electromagnetic spectrum, and the range of PBH masses $M$ for which $E_{\rm peak}$ for photons from Hawking radiation falls within a given spectral band. Note that peak emission from asteroid-mass PBHs spans from the infrared to the $\gamma$-ray bands.
}
\begin{ruledtabular}
\begin{tabular}{ccc}
 Band  & Peak primary photon energy $E_{\rm peak}$ [eV] & PBH mass [g] \\
\hline
infrared & $1.24 \times 10^{-3} \leq E_{\rm peak} \leq 1.7$ & $3.8 \times 10^{22} \leq M \leq 5.1 \times 10^{25}$ \\
visible & $1.7 \leq E_{\rm peak} \leq 3.3$ & $1.9 \times 10^{22} \leq M \leq 3.8\times 10^{22}$ \\
ultraviolet & $3.3 \leq E_{\rm peak} \leq 124$ & $5.1 \times 10^{20} \leq M \leq 1.9 \times 10^{22}$ \\
X-ray & $124 \leq E_{\rm peak} \leq 1.24 \times 10^5$ & $5.1 \times 10^{17} \leq M \leq 5.1 \times 10^{20}$ \\
$\gamma$-ray & $E_{\rm peak} > 1.24 \times 10^5$ & $ M \leq 5.1\times 10^{17}$ \\
\end{tabular}
\end{ruledtabular}
\end{table*}

In Section ~\ref{sec:Hawking}, we review the primary and secondary Hawking emission formalism, which is well-studied in the literature. In Section~\ref{sec:Transits} we present methods to simulate the time-varying photon signal measured by a detector at Earth during an inner Solar System PBH transit. We consider prospects for measuring EM signals from PBH transits in the X-ray, ultraviolet (UV), and radio bands and compute the maximum detectable impact parameter as a function of PBH mass for instruments sensitive to each band.
In Section~\ref{sec:Explosions}, we study the prospects for detecting photons from local PBH explosions. 
Many prior studies by ground- and space-based cosmic ray observatories constrain the local PBH burst rate in the neighborhood of the Solar System~\cite{Linton:2006yu,Alexandreas:1993zx,Glicenstein:2013vha, Archambault:2017asc,Fermi-LAT:2018pfs,Abdo:2014apa,HAWC:2019wla,HESS:2023zzd, LHAASO:2025kyn}, and analysis of realistic PBH number distributions allows direct calculation of expected burst rates for a given PBH population~\cite{Klipfel:2025jql}. We specifically evaluate the feasibility of detecting a PBH explosion 
in the outskirts of the Solar System and discuss multimessenger detection prospects for ultrahigh-energy photons and neutrinos.

We work in so-called ``natural units," in which we set $c = \hbar = k_B = 1$. In these units, all dimensionful quantities can be represented as either a {\it mass} $[M]$ or a {\it length} $[ L ] \sim [ M^{-1} ]$. The fundamental unit of electric charge $e$ is dimensionless in these units (and the dielectric constant $\epsilon_0 = 1$), with value $e = 0.303$. Newton's gravitational constant $G$ may be represented in terms of the reduced Planck mass, $M_{\rm pl} \equiv 1 / \sqrt{ 8 \pi G} = 2.43 \times 10^{18} \, {\rm GeV} = 4.33 \times 10^{-6} \, {\rm g}$.

\section{Hawking Emission of Photons}
\label{sec:Hawking}

Black holes couple to every form of matter, due to the universal nature of gravitation. Hence PBHs of sufficiently high temperature
can radiate every fundamental particle. 
Whether and how the semi-classical Hawking-radiation formalism might need to be modified at late stages of black hole evaporation remains an open question~\cite{Dvali:2018xpy,Dvali:2020wft,Zantedeschi:2024ram,Dvali:2025ktz,Montefalcone:2025akm}. As a conservative analysis, we work with the standard formalism~\cite{Hawking:1974rv,Hawking:1975vcx,Page:1976df,Page:1976ki,Page:1977um,macgibbonQuarkGluonjetEmission1990, macgibbonQuarkGluonjetEmission1991}, updated as in Refs.~\cite{Klipfel:2025jql,Klipfel:2025bvh,Klipfel:2026aug} to include the present-day set of Standard Model (SM) degrees of freedom.

We assume the PBHs of interest have no charge and no spin. PBHs form from the collapse of (scalar) curvature perturbations, so they typically begin with little or no spin~\cite{Garc_a_Bellido_2017}.  Similarly, although small-mass PBHs could have formed with large initial charge~\cite{Alonso-Monsalve:2023brx}, such short-lived objects are expected to discharge very rapidly (if charged only under SM gauge groups ~\cite{Baker:2025zxm,Santiago:2025rzb}). Meanwhile, following their formation, black holes that happened to form with charge and/or spin will preferentially emit particles to reduce those quantities~\cite{carterChargeParticleConservation1974,gibbonsVacuumPolarizationSpontaneous1975,Page:1976df}. Furthermore, PBHs within the mass range we consider here will not spin up over time: their extraordinarily small radii preclude efficient accretion~\cite{chibaSpinDistributionPrimordial2017,delucaEvolutionPrimordialBlack2020,Jaraba:2021ces,chongchitnanExtremeValueStatisticsSpin2021}. We therefore focus on emission from Schwarzschild PBHs and their evaporation over time. (If the emitting PBH \emph{did} have nonzero angular momentum, one could aim to measure its spin by careful analysis of the angular dependence of its neutrino and photon Hawking emission \cite{Perez-Gonzalez:2023uoi}.) 

\subsection{Primary Spectra }

The primary spectrum of Hawking emission may be parameterized as
\begin{equation}
    \frac{ d^2 N_j^{(1)}}{dt dE}  = g_j \frac{ \Gamma_{s_j}}{2 \pi} \left[ \exp \left( \frac{ E}{T_H}\right) - (-1)^{2 s_j} \right]^{-1}.
\label{Hawk1}
\end{equation}
The subscript $j$ refers to particle species, $g_j$ is the number of degrees of freedom associated with that species, $s_j$ is the particle's spin, $\Gamma_{s_j}$ is the greybody factor, $E$ is the energy of the emitted particle, and 
\begin{equation}
    T_H = \frac{1}{8 \pi G M} 
\label{TH}
\end{equation}
is the Hawking temperature of the black hole.

\begin{figure*}[t]
    \includegraphics[width=0.49\textwidth]{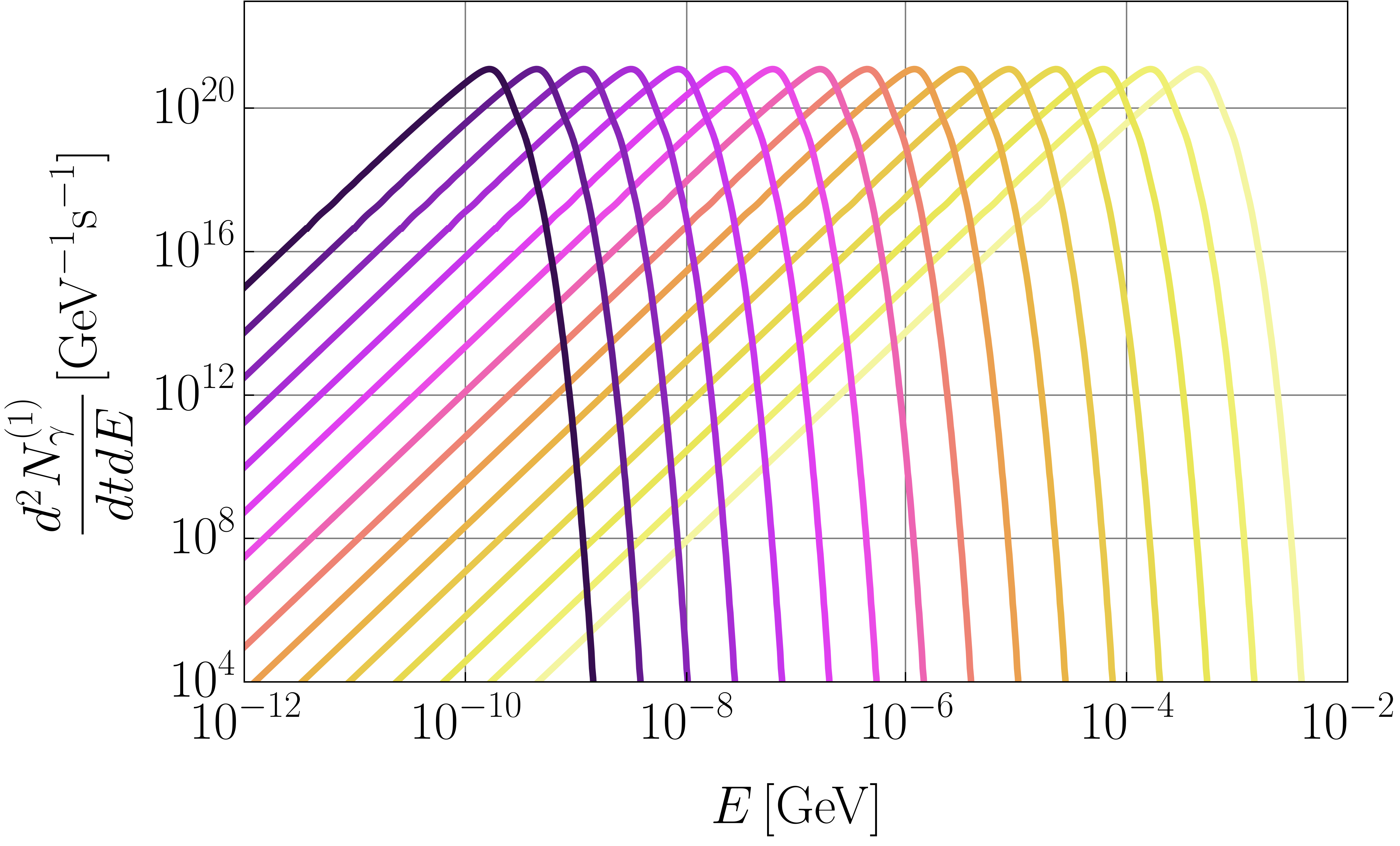} 
    \includegraphics[width=0.49\textwidth]{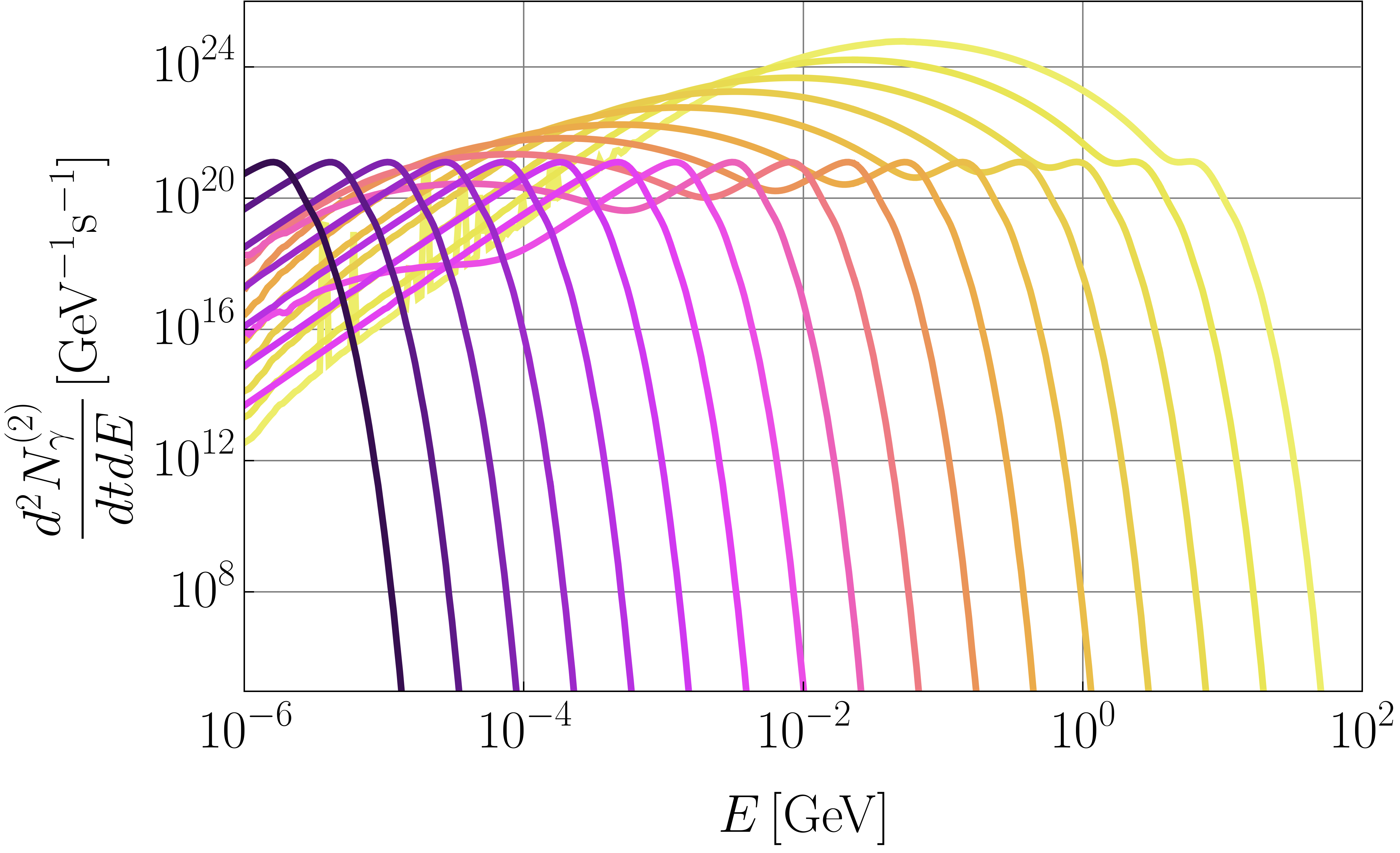}
    \caption{\justifying (\emph{Left}) Primary photon Hawking emission spectra for Schwarzschild black holes with masses ranging from $10^{17} \, {\rm g}$ (yellow) to $10^{24} \, {\rm g}$ (black).  (\emph{Right}) Secondary Hawking emission spectra for PBHs with masses ranging from $10^{13}\,{\rm g}$ (yellow) to $10^{20}\,{\rm g}$ (black). Plots prepared with \texttt{BlackHawk v2.2} \cite{arbeyBlackHawkV20Public2019, arbeyPhysicsStandardModel2021a}. }
    \label{fig:HawkPrimary}
\end{figure*}

For the case of interest---namely, emission of photons---we have $g_j = 2$ for the number of distinct polarization states, and $s_j = 1$ for the photon's spin. The greybody factor for photons may be parameterized as \cite{macgibbonQuarkGluonjetEmission1990}
\begin{equation}
    \Gamma_\gamma = \frac{ \sigma_s (E, M)}{ \pi} E^2 .
\label{greybody1}
\end{equation}
Note that the greybody factor is dimensionless. 
The cross sections $\sigma_s (E, M)$ must be evaluated numerically for arbitrary $E$; only their analytic forms for $E \rightarrow 0$ and $E \rightarrow \infty$ are known \cite{pageParticleEmissionRates1976}. In general the values of $\sigma_s (E, M)$ arise from solving for the transmission coefficients of modes scattering in the Regge-Wheeler effective potential. (See Ref.~\cite{grayGreybodyFactorsSchwarzschild2018} for a helpful discussion.) To calculate the full spectrum of all emitted particles, one may use \texttt{BlackHawk} \cite{arbeyBlackHawkV20Public2019,arbeyPhysicsStandardModel2021a}.  Primary photon emission spectra are shown in Fig.~\ref{fig:HawkPrimary}.

In Fig.~\ref{fig:HawkPrimary} we can see that the primary photon emission spectra are highly peaked. The photon energy at the peak of the distribution is given by \cite{macgibbonQuarkGluonjetEmission1990, macgibbonQuarkGluonjetEmission1991}
\begin{equation}
    E_{\rm peak} \approx 6.04 \, T_H ,
\label{EpeakHawk}
\end{equation}
where $T_H$ is the Hawking temperature as given in Eq.~(\ref{TH}). The integrated primary photon emission rate obeys
\begin{equation}
    \frac{ dN_\gamma^{(1)}}{dt}  = \int_0^\infty dE \frac{ d^2 N_\gamma^{(1)}}{dtdE}  = 5.97 \times 10^{17} \, {\rm s}^{-1} \left( \frac{10^{17} \, {\rm g}}{ M } \right) ,
\label{dNdtHawk}
\end{equation}
scaling inversely with $M$. 

As expected, smaller-mass PBHs are hotter and hence emit more primary photons, with a greater typical energy, than larger-mass PBHs. Note that $E_{\rm peak}$ ranges broadly across the asteroid-mass range, from $E_{\rm peak} \sim {\cal O} (10^{-1} \, {\rm eV})$ for $M \sim 10^{23} \, {\rm g}$ to $E_{\rm peak} \sim {\cal O} (10^5 \, {\rm eV})$ for $M \sim 10^{17} \, {\rm g}$. In Table~\ref{tab:HawkIRUV}, we relate the PBH mass $M$ to the portion of the electromagnetic spectrum in which $E_{\rm peak}$ resides across the asteroid-mass range. As indicated there, Hawking emission of photons from PBHs within the asteroid-mass range spans all the way from the infrared to the gamma-ray bands.

\subsection{Secondary Spectra}

We use \texttt{BlackHawk~v2.2} \cite{arbeyBlackHawkV20Public2019, arbeyPhysicsStandardModel2021a}, \texttt{PYTHIA} \cite{sjostrandIntroductionPYTHIA822015},
and \texttt{HDMSpectra} \cite{bauerDarkMatterSpectra2021} to compute the \textit{secondary} Hawking emission spectrum for photons, which is defined as:
\begin{equation}
    \label{HawkSecondary}
    \frac{d^2N_{\gamma}^{(2)}}{dtdE} = \int_0^{\infty} \sum_j \frac{d^2N^{(1)}_{j}}{dtdE'} \frac{dN_{\gamma}^j}{dE}dE',
\end{equation}
where $dN_{\gamma}^j(E',E)/dE$ are the differential branching ratios \cite{arbeyBlackHawkV20Public2019}. The secondary photon spectra for PBHs with masses $10^{13}\,{\rm g}\leq M \leq10^{20}\,{\rm g}$ are plotted in Fig.~\ref{fig:HawkPrimary}. 
For PBHs with $M\gtrsim5\times10^{17}\,{\rm g}$, electron-positron emission is negligible and the secondary photon spectrum converges to the primary photon spectrum.

\section{Signatures from a Single PBH Transit}
\label{sec:Transits}

In this section we focus on strategies to detect stable, ``quiescent'' PBHs that may transit through the inner Solar System. Such PBHs would have masses $M > 5\times10^{14}\,{\rm g}$ and lifetimes exponentially larger than the current age of the universe. PBH transits within $\mathcal{O}(5\,{\rm AU})$ of the Earth could produce measurable perturbations to the orbit of Mars~\cite{tranCloseEncountersPrimordial2024}, for which the detection of Hawking-radiated photons would be a valuable multimessenger signature to differentiate a PBH from a mundane asteroid perturber. We expand on previous work estimating photon Hawking radiation signals from  Ref.~\cite{sobrinhoDirectDetectionBlack2014} by using numerical Hawking spectra computed with \texttt{BlackHawk v2.2}~\cite{arbeyBlackHawkV20Public2019, arbeyPhysicsStandardModel2021a} and by applying the methods of Ref.~\cite{Klipfel:2025bvh}, which take into account the relative motion between the PBH and the detector to simulate time-dependent photon signals for specific detector geometries.

Many constraints on the sub-asteroid-mass PBH dark matter fraction
are derived from galactic and extragalactic gamma-ray observations, including extra-galactic gamma ray bursts \cite{carrNewCosmologicalConstraints2010b}, galactic gamma-ray bursts \cite{carrConstraintsPrimordialBlack2016}, galactic MeV diffuse flux \cite{lahaINTEGRALConstraintsPrimordial2020}, MeV spectrum from nearby galaxies \cite{cooganDirectDetectionHawking2021}, the galactic 511 keV line \cite{deroccoConstrainingPrimordialBlack2019, lahaPrimordialBlackHoles2019a}, and a Fermi-LAT search for local PBH explosions \cite{thefermi-latcollaborationSearchGammaRayEmission2018}. The furthest that any of these monochromatic constraints push into the asteroid mass window is up to $M\simeq 2\times10^{17} \, {\rm g}$ with INTEGRAL observations \cite{lahaINTEGRALConstraintsPrimordial2020,DelaTorreLuque:2024qms,Balaji:2025afr}. 
As shown in Table \ref{tab:HawkIRUV}, however, low-energy (sub-MeV) datasets are required to constrain Hawking emission from PBHs with masses $M\gtrsim 5\times10^{17} \, {\rm g}$. We note that lower-energy photon signals from a PBH population throughout the galaxy are typically complicated to model because of propagation effects such as energy loss and scattering off the interstellar medium (ISM). However, we can neglect these phenomena (which would attenuate the signal) for inner Solar System transits because of the short distance-scales involved ($\sim 5\, {\rm AU}$) and the diffuse nature of the interplanetary medium (IPM). Thus, searching for the time-dependent signatures of individual PBH transits in the inner Solar System, rather than galactic photon fluxes, avoids model-dependencies from photon propagation and the galactic composition.

In this section we model photon signals from asteroid-mass and sub-asteroid-mass PBHs transiting through the inner Solar System and compare signal strength to expected background in various energy bands in which existing or proposed experiments have sensitivity. 
In Section~\ref{sec:TRates}, we first generalize monochromatic constraints on $f_{\rm PBH}$ to extended mass functions and then estimate expected PBH transit rates through the inner Solar System. Then in Section~\ref{sec:PhotonSignals} we apply the methods of Ref.~\cite{Klipfel:2025bvh} to simulate measured photon signals from a PBH transit for a given detector geometry. In Section~\ref{sec:SpaceBased}, we evaluate prospects for PBH detection with GALEX and the proposed AMEGO-X instrument, which are space-based detectors sensitive to UV and X-ray bands, respectively. Finally, in Section~\ref{sec:Radio} we briefly address prospects for detecting PBH emission in the radio band.

\subsection{PBH Transit Rates}
\label{sec:TRates}

The PBH number distribution function at time of formation $t_i$ is defined via:
\begin{equation}
    \phi(M_i) = \frac{1}{n_{{\rm PBH},i}}\frac{dn_{{\rm PBH},i}}{dM_i},
\end{equation}
where $M_i$ is the initial PBH mass at formation and $n_{{\rm PBH},i}$ the initial PBH number density. PBH populations which form from critical collapse of a radiation fluid in the aftermath of inflation are expected to obey a number distribution function with the form~\cite{gorton_how_2024,Mosbech:2022lfg,Klipfel:2025bvh}:
\begin{equation}
    \label{eqn:PhiGCC}
    \begin{split}
    \phi_{\rm GCC}&(M_i) \propto \\
    & \frac{1}{\bar{M}} \left( \frac{M_i}{\bar{M}}\right)^{\alpha-1}\exp\left[-\left(\frac{\alpha-1}{\beta} \right)\left(\frac{M_i}{\bar{M}} \right)^{\beta}\right],
    \end{split}
\end{equation}
which is peaked at mass $\bar{M}$. The constant $\alpha > 1$ controls the small-mass power-law tail for $M_i < \bar{M}$, while $\beta > 0$ controls the (super)-exponential cutoff for $M_i > \bar{M}.$ 
For PBHs that form from the collapse of a Gaussian spectrum of primordial curvature perturbations, we expect $\alpha=\beta=2.78$, the so-called ``critical collapse'' parameter values \cite{Carr:2020xqk,greenPrimordialBlackHoles2021,Escriva:2022duf}. The subscript ``GCC'' refers to ``generalized critical collapse'' distributions.

For PBHs that form with masses near the asteroid-mass range ($M_i \ll M_\odot$), accretion remains negligible even over cosmological time-scales, due to the PBHs' sub-micron Schwarzschild radii, $r_s = 2 G M$ \cite{Rice:2017avg,Alonso-Monsalve:2023brx}. On the other hand, the mass of individual PBHs will change for times $t > t_i$ due to mass-loss via Hawking emission, at a rate
\cite{macgibbonQuarkGluonjetEmission1990, macgibbonQuarkGluonjetEmission1991}
\begin{equation}
    \frac{dM}{dt}=-{\cal A}\frac{F(M)}{M^2}.
\end{equation}
The \emph{Page factor} $F(M)$ quantifies the number of degrees of freedom which can be emitted by a PBH of mass $M$.\footnote{We denote the Page factor as $F(M)$ here, rather than $f(M)$ as in our recent papers \cite{Klipfel:2025jql,Klipfel:2025bvh,Klipfel:2026aug,Vanvlasselaer:2026vkh,Klipfel:2026nzx}, to avoid confusion with $f_{\rm PBH}$, the PBH fraction of the dark matter density.} Using the Page factor parameterization from Refs.~\cite{macgibbonQuarkGluonjetEmission1990, macgibbonQuarkGluonjetEmission1991,Vanvlasselaer:2026vkh,Klipfel:2026nzx} with modern particle values from Ref.~\cite{Klipfel:2025jql} and normalizing $F(M)$ such that $F(M)=1$ for PBHs that emit only photons and neutrinos (assumed massless), we find that the Page factor is a function which smoothly interpolates between $F_{\rm min}\equiv F(M\gg M_*)=1$ at large masses ($M \gtrsim 10^{18} \, {\rm g}$) and $F_{\rm max}\equiv F(M\ll M_*)=15.522$ at small masses ($M \lesssim 10^9 \, {\rm g}$). A PBH that forms with mass $M_i = M_*$, where
\begin{equation}
    M_* = 5.364\times10^{14}\,{\rm g} ,
\label{Mstar}
\end{equation}
has a lifetime equal to the current age of the universe, $t_0= 13.787 \,{\rm Gyr}$ \cite{Klipfel:2025bvh}. Given this parameterization of the Page factor, we can take the constant to be ${\cal A}=5.195\times10^{25}\,{\rm g}^3{\rm s}^{-1}$ \cite{Vanvlasselaer:2026vkh}.

The number distribution function will evolve in time to some $\phi(M, t)$ for $t>t_i$ as the PBHs radiate and eventually explode \cite{Klipfel:2025bvh, Mosbech:2022lfg, Klipfel:2025jql}. 
For mass distributions with $\bar{M}\gg M_*$, the time-evolved number distribution takes the form \cite{Klipfel:2025bvh}:
\begin{equation}
    \begin{split}
    \phi(M, t) & \simeq \frac{M^2 \,\phi_{\rm GCC}(M_i(M,t))}{\left[M^3 + 3A F_{\rm min}t \right]^{2/3}},
    \end{split}
    \label{eq:phiEvol}
\end{equation}
where 
the mass relationship is approximated by
\begin{equation}
    M_i(M,t)\simeq(M^3+3 {\cal A} F_{\rm min}t)^{1/3}.
\end{equation}
Note that we normalize the \emph{present-day} number distribution function such that $\int dM\,\phi(M, t_0)=1$.

For the distributions of interest with $\bar{M}\gg M_*$, the majority of the PBH population is in a regime of low Hawking emission rates, with lifetimes exponentially longer than the present age of the universe, so the peak of the distribution will not vary significantly over cosmological time-scales of order $t_0$ \cite{Klipfel:2025bvh}. Furthermore, these PBHs in the bulk of the population ($M\simeq \bar{M}$) have sufficiently low Hawking temperatures to only emit photons and neutrinos---which is why the population evolution is best modeled by approximating the Page factor by its minimum value $F_{\rm min}$. We note that this choice of Page factor results in a less accurate model of the small-mass tail of the present-day distribution $\phi(M,t_0)$, which does not contribute significantly to the transit rate.

The PBH mass density $\rho_{ {\rm PBH}}$ is related to the number distribution function $\phi$ and the PBH number density $n_{{\rm PBH}}$ at the present day $t_0$ by
\begin{equation}
    \rho_{\rm PBH}(t_0)= n_{\rm PBH}(t_0)\int_0^{\infty} dM \,M\, \phi(M,t_0).
    \label{eq:rhon1}
\end{equation}
We define the constant $\rho_{\rm PBH}(t_0)$ in terms of the present-day local dark matter density and the PBH dark matter fraction $f_{\rm PBH}$: 
\begin{equation}
    \rho_{\rm PBH}=f_{\rm PBH}(\bar{M}, \alpha, \beta)\, \rho_{\rm DM}^{\odot},
    \label{eq:rhoPBHdef}
\end{equation}
where we take $\rho_{\rm DM}^{\odot}=0.0155 \,M_{\odot}{\rm pc}^{-3}$ \cite{Klipfel:2025bvh}. As in Refs.~\cite{Klipfel:2025jql,Klipfel:2025bvh}, we assume that the PBHs are distributed throughout the Milky Way galaxy in a way that tracks a modified Navarro-Frenk-White (NFW) dark matter density profile, and we neglect possible effects of PBH clustering \cite{carrObservationalEvidencePrimordial2024,Escriva:2022duf}. We further assume that PBHs and any other contributions to the local dark matter density are distributed within the Milky Way in the same relative proportions as throughout the universe, and that this ratio has not changed significantly over cosmological time-scales. Hence the parameter $f_{\rm PBH}$ that we introduce in Eq.~(\ref{eq:rhoPBHdef}), which is defined in terms of the present-day local PBH fraction of the local dark matter density, should be compatible with definitions of $f_{\rm PBH}$ defined in terms of the initial (primordial) fraction PBHs of the total dark matter density.

\begin{figure}[t]
    \includegraphics[width=0.49\textwidth]{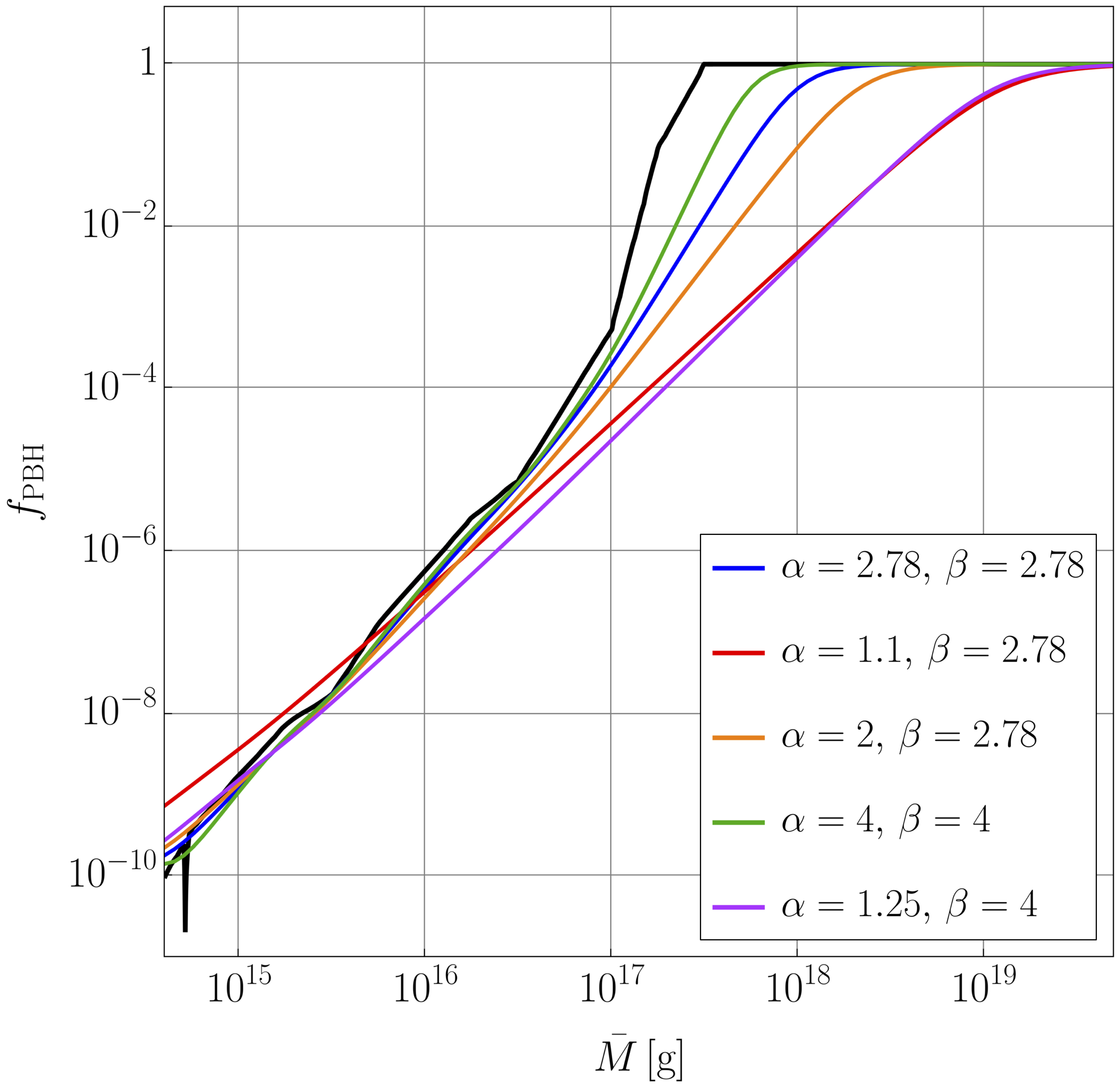} 
    \caption{\justifying Constraints on PBH dark matter fraction $f_{\rm PBH}$ for extended GCC mass functions peaked at $\bar{M}$ with model parameters $\alpha$ and $\beta$, as computed with Eq.~(\ref{eq:fPBHemf}). The black curve consists of evaporation constraints on monochromatic distributions from Refs.~\cite{carrConstraintsPrimordialBlack2021,Balaji:2025afr}.}
    \label{fig:fPBH}
\end{figure}

\begin{figure*}[t]
    \includegraphics[width=0.49\textwidth]{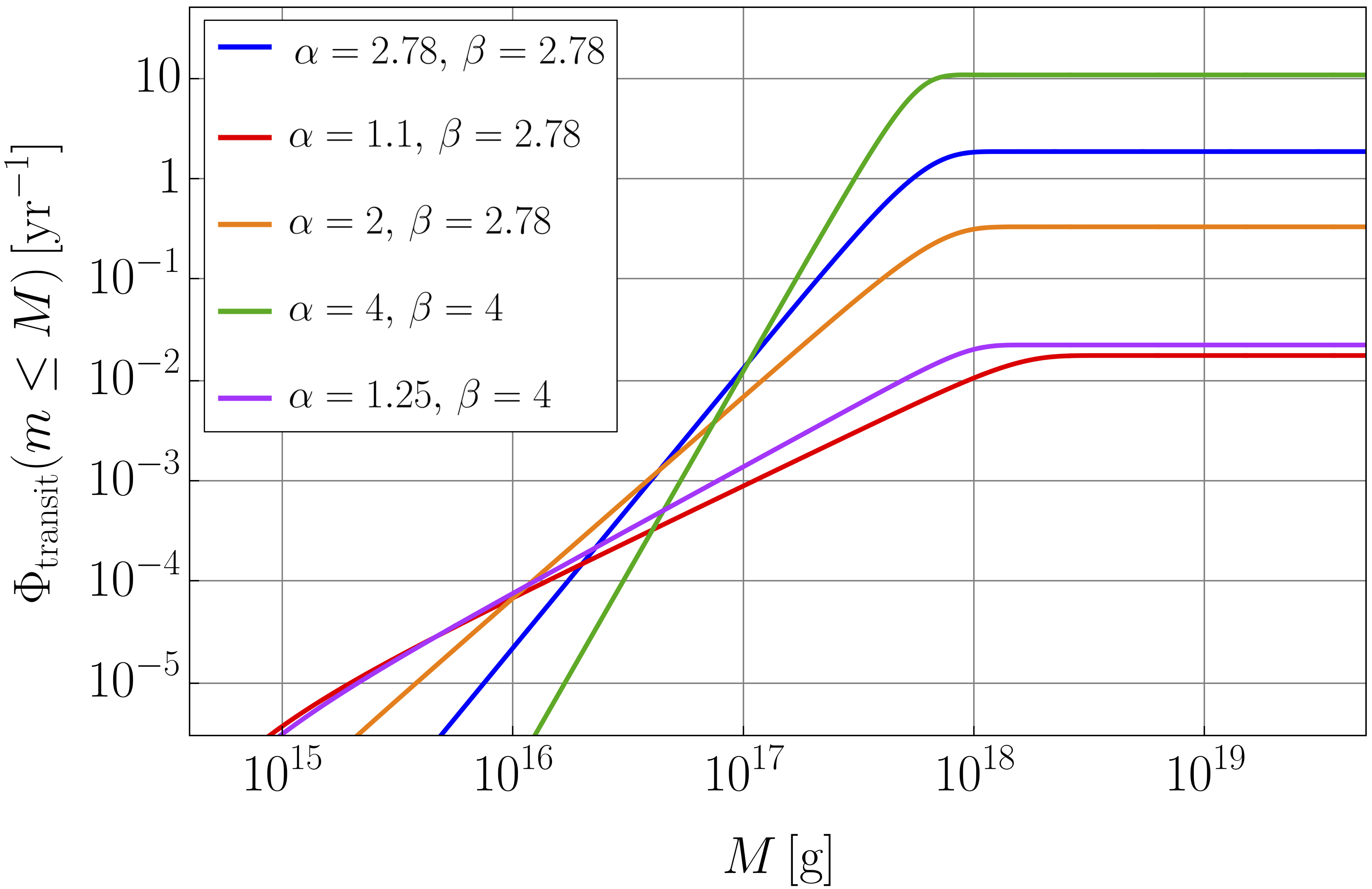} 
    \hfill
    \includegraphics[width=0.49\textwidth]{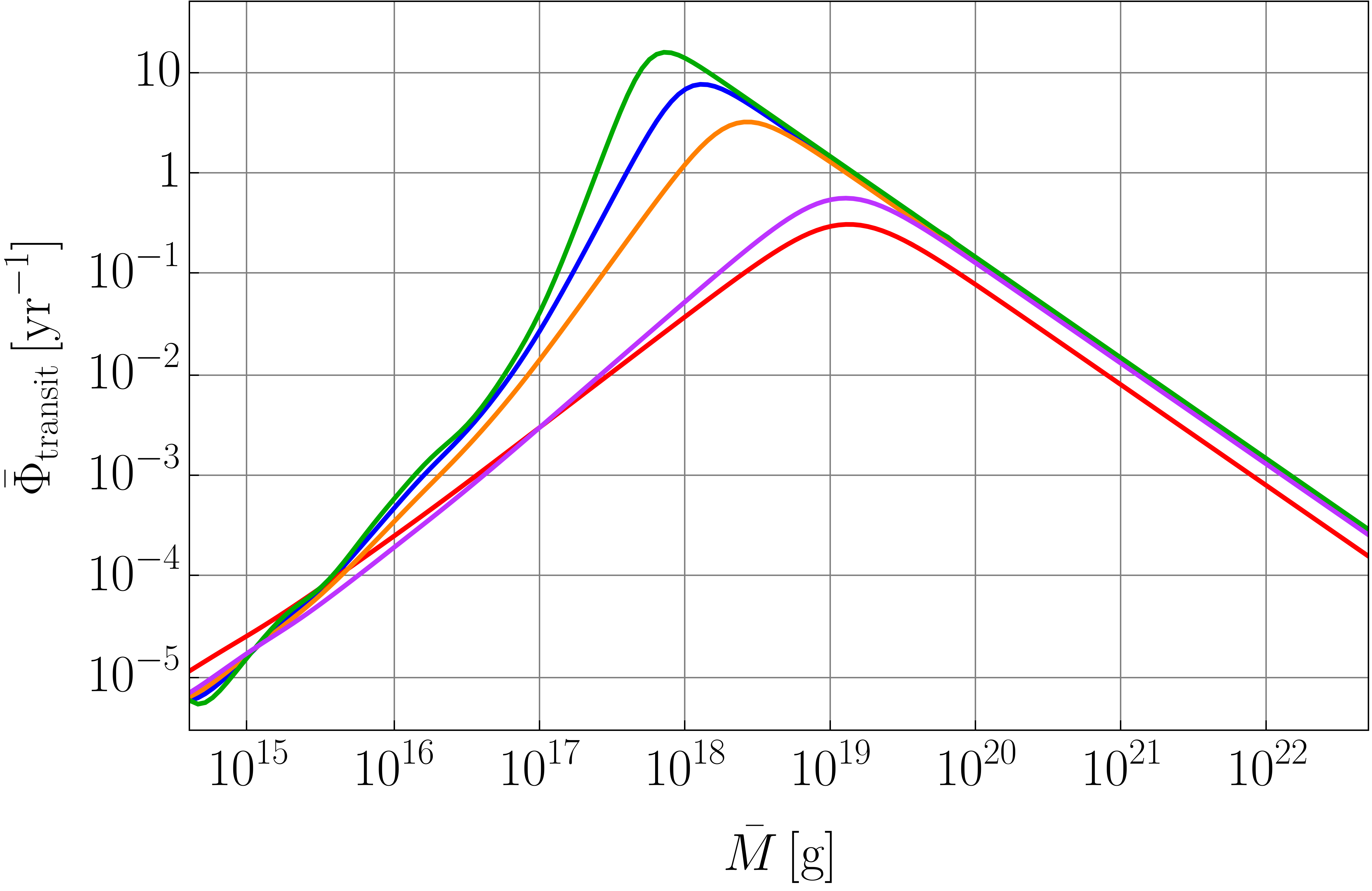}
    \caption{\justifying (\emph{Left}) Cumulative inner Solar System transit rates for PBHs with mass $m\leq M$ expected to pass within $5\,{\rm AU}$ of the Earth per year, according to Eq.~(\ref{eq:cumulativePhi}). The peak of the number distribution is fixed at $\bar{M}=5\times10^{17}\,{\rm g}$. (\emph{Right}) Total PBH transit rates for impact parameters $b\leq 5 \, {\rm AU}$ as a function of population peak mass $\bar{M}$ and model parameters $\alpha$ and $\beta$, computed with Eq.~(\ref{eq:avgPhi}). We consider transit rates for the same set of representative GCC mass functions as analyzed in Fig.~\ref{fig:fPBH}. Note that larger values of $\alpha$ and $\beta$ correspond to more sharply-peaked mass functions, so results approach the monochromatic limit for $\alpha=\beta\gtrsim4$.}
    \label{fig:TRates}
\end{figure*}

We compute the PBH dark matter fraction for an \emph{extended} mass distribution with parameters $\bar{M}, \alpha, \beta$ from published constraints on \emph{monochromatic} distributions by following the methods of Refs.~\cite{gorton_how_2024, carrPrimordialBlackHole2017}. Given the upper limit constraints on $f_{\rm PBH}$ for a monochromatic distribution peaked at mass $M$, which we call $f_{\rm mono}(M)$, we can compute the maximum allowed value for $f_{\rm PBH}$ given an extended mass distribution via
\begin{equation}
    f_{\rm PBH}(\bar{M}, \alpha, \beta)\leq \left[\int_0^\infty dM \frac{\psi(M, t_0|\bar{M}, \alpha, \beta)}{f_{\rm mono}(M)} \right]^{-1},
    \label{eq:fPBHemf}
\end{equation}
where the normalized \emph{mass function} $\psi$ is defined by:
\begin{equation}
    \psi(M, t_0|\bar{M},\alpha, \beta)=\frac{M\,\phi(M,t_0|\bar{M},\alpha, \beta)}{\int dM\,M\,\phi(M,t_0|\bar{M},\alpha, \beta)}.
\end{equation}
We take $f_{\rm mono}(M)$ from Ref.~\cite{carrConstraintsPrimordialBlack2021} and Ref.~\cite{Balaji:2025afr}, which places the tightest constraints on the lower bound of the asteroid-mass window with INTEGRAL/SPI observations.

We can now combine Eqs.~(\ref{eq:rhon1}) and (\ref{eq:rhoPBHdef}) to express the present-day total PBH number density in terms of known quantities and model parameters:
\begin{equation}
    n_{\rm PBH}(t_0) = \frac{f_{\rm PBH}(\bar{M},\alpha,\beta)\,\rho_{\rm DM}^\odot}{\int_0^\infty dM \,M \,\phi(M,t_0|\bar{M},\alpha,\beta)}.
\end{equation}

The expected transit rate within some impact parameter $b$ of the Earth for PBHs with mass $m\leq M$ is then given by
\begin{equation}
    \begin{split}
    \Phi_{\rm transit}&(b, m\leq M)\\
    &=\pi b^2 \bar{v}\,n_{\rm PBH}(t_0)\int_0^M dM'\,\phi(M',t_0)
    \end{split}
    \label{eq:cumulativePhi}
\end{equation}
where $\bar{v}=246\,{\rm km/s}$ is the average relative PBH velocity \cite{Klipfel:2025bvh}.
The total transit rate within distance $b$ of the Earth for PBHs of all masses is thus
\begin{equation}
    \bar{\Phi}_{\rm transit}(b|\bar{M},\alpha,\beta)=\frac{\pi b^2 \bar{v}\,f_{\rm PBH}(\bar{M},\alpha,\beta)\,\rho_{\rm DM}^\odot}{\int_0^\infty dM \,M \,\phi(M,t_0|\bar{M},\alpha,\beta)}
    \label{eq:avgPhi}
\end{equation}
Figure~\ref{fig:TRates} plots expected PBH transit rates within $b\leq5\,{\rm AU}$ of Earth for several representative number distribution functions. We note that for $10^{17}\,{\rm g} \leq \bar{M}\leq10^{21}\,{\rm g}$, the transit rates vary from about one per century to several dozen per year, implying that detection may be possible on typical human and experimental time-scales.

\subsection{Time-Dependent Photon Signals from PBH Transits}
\label{sec:PhotonSignals}

In this section, we build upon the methods developed in Ref.~\cite{Klipfel:2025bvh} to compute time-dependent photon signals from a local PBH transit as measured by a detector in orbit about Earth or fixed to the surface of a rotating Earth. For a given instrument, we consider a sensitive energy band bounded by minimum and maximum detectable photon energies $[E_{\rm min}, \, E_{\rm max}]$. We assume the instrument has some energy-dependent detection efficiency $\epsilon(E) \equiv A_{\rm eff}(E)/A_{\rm geo}$, where $A_{\rm geo}$
is the physical geometric area of the detector. In this section we give generic results, applicable to any detector geometry, and then specialize to specific instruments in Section~\ref{sec:SpaceBased}. 

The emission rate of \textit{detectable} photons for a PBH of mass $M$ is: 
\begin{equation}
    \label{DethawkPhotons}
    \left(\frac{dN_{\gamma}}{dt}\right)_{\rm det} = \int_{E_{\rm min}}^{E_{\rm max}}dE \,\frac{d^2N_{\gamma}^{(2)}}{dtdE} \epsilon(E).
\end{equation}
See Fig.~\ref{fig:AXdetrate} for a plot of $(dN_\gamma/dt)_{\rm det}$ for the proposed AMEGO-X experiment, with sensitive energy range $10\,{\rm keV}\leq E \leq 1 \, {\rm GeV}$.
The secondary Hawking spectra are computed numerically via \texttt{BlackHawk v2.2}~\cite{arbeyBlackHawkV20Public2019, arbeyPhysicsStandardModel2021a} for a given PBH mass.

We consider general PBH trajectories past Earth with velocity $v$ sampled from a Maxwellian distribution \cite{Klipfel:2025bvh}:
\begin{equation}
    \label{eqn:Maxwellian}
    f(v) = \frac{4 f_0}{\sqrt{\pi}} \left( \frac{3}{2}\right)^{3/2} \frac{v^2}{v_{\text{rms}}^3} \exp\left[ - \frac{3}{2}\frac{v^2}{v_{\text{rms}}^2} \right]; \ \ v < v_{\text{esc}}.
\end{equation}
We assume that the Sun is located at $r_{\odot} = 8.0\pm0.5$ kpc in Galactocentric coordinates, which corresponds to $v_{\odot}=220\pm20$ km/s \cite{cerdenoParticleDarkMatter2010}. The velocity dispersion is related to the Sun's azimuthal velocity in the galactic plane via $v_{\text{rms}} = \sqrt{3/2}\, v_{\odot} \approx 270$ km/s \cite{choiImpactDarkMatter2014}. Truncating Eq.~(\ref{eqn:Maxwellian}) at the galactic escape velocity $v_{\text{esc}}=544$ km/s sets the normalization constant $f_0=1.00668$.

We want to generate a random linear PBH trajectory with some specified distance of closest approach, or \emph{impact parameter}, $b$. The set of all lines in $\mathbb{R}^3$ with only one point a distance $b$ from the origin is exactly the set of all tangent lines to a sphere of radius $b$ centered at the origin. Thus, generating a random PBH trajectory amounts to sampling a random tangent line to a sphere of radius $b$, which can be accomplished via the following procedure.

Sample a random point on a sphere of radius $b$: 
\begin{equation}
    \vec{r}_1(b, \theta_1, \phi_1) = b\, \langle\sin{\theta_1}\sin{\phi_1}, \, \sin{\theta_1}\cos{\phi_1}, \, \cos{\theta_1}\rangle,
\end{equation}
where $\theta_1 \in [0, \pi]$ and $\phi_1 \in [0, 2\pi)$. Then construct a basis for the tangent plane at $\vec{r}_1$:
\begin{equation}
    \begin{split}
        & \vec{e}_1  = \frac{\partial_{\theta}\vec{r}_1(b, \theta, \phi_1)}{|\partial_{\theta}\vec{r}_1(b, \theta, \phi_1)|} \Bigg|_{\theta = \theta_1} \\
        & \vec{e}_2 = \frac{\vec{r}_1(b, \theta_1, \phi_1) \times \vec{e}_1}{|\vec{r}_1(b, \theta_1, \phi_1) \times \vec{e}_1|}.
    \end{split}
\end{equation}
A random vector in this plane can then be specified by a randomly sampled phase $\psi_2 \in [0, 2\pi)$:
\begin{equation}
    \vec{r}_2 = \cos{(\psi_2)}\, \vec{e}_1 + \sin{(\psi_2)}\, \vec{e}_2.
\end{equation}
Finally, we construct two points in the Earth-centered coordinate system that define a randomly sampled tangent line: $\{\vec{r}_1, \vec{r}_1 + \vec{r}_2 \}$. 

\begin{table*}[t]
\caption{\label{tab:ZetaParams} \justifying Symbol definitions and values for PBH transit simulation parameters and relevant variables. The time-dependent measured photon count rate $\zeta_{\gamma}(t)$ given by Eq.~(\ref{MeasPhotonCountRate}) depends on all 12 parameters listed here.
}
\begin{ruledtabular}
\begin{tabular}{llll}
Symbol & Definition   & Units \\ 
\hline
$M$         & PBH mass           & g \\
$v$         & PBH velocity sampled from $f(v)$ in Eq.~(\ref{eqn:Maxwellian})     & m/s \\
$t$         & Time since closest approach & s\\
$h$         & Orbit altitude       & km \\
$\tau$      & Orbit period       & s\\
$\theta_{\rm FOV}$    & Detector field of view angle          & radians \\
$A$& Detector area   & m$^2$\\     
$E_{\rm min}$   & Detector minimum detectable photon energy & GeV\\
$E_{\rm max}$   & Detector maximum detectable photon energy & GeV\\
$\theta_{\rm inc}$   & Detector orbit inclination angle relative to equatorial plane & radians\\
$\epsilon(E)$   & Energy-dependent detection efficiency  &  \\
$\vec{r}_j, \, j=1,2$   & Points that define the PBH trajectory path & \\
\end{tabular}
\end{ruledtabular}
\end{table*}

\begin{figure}[t]
    \centering
    \includegraphics[width=0.85\linewidth]{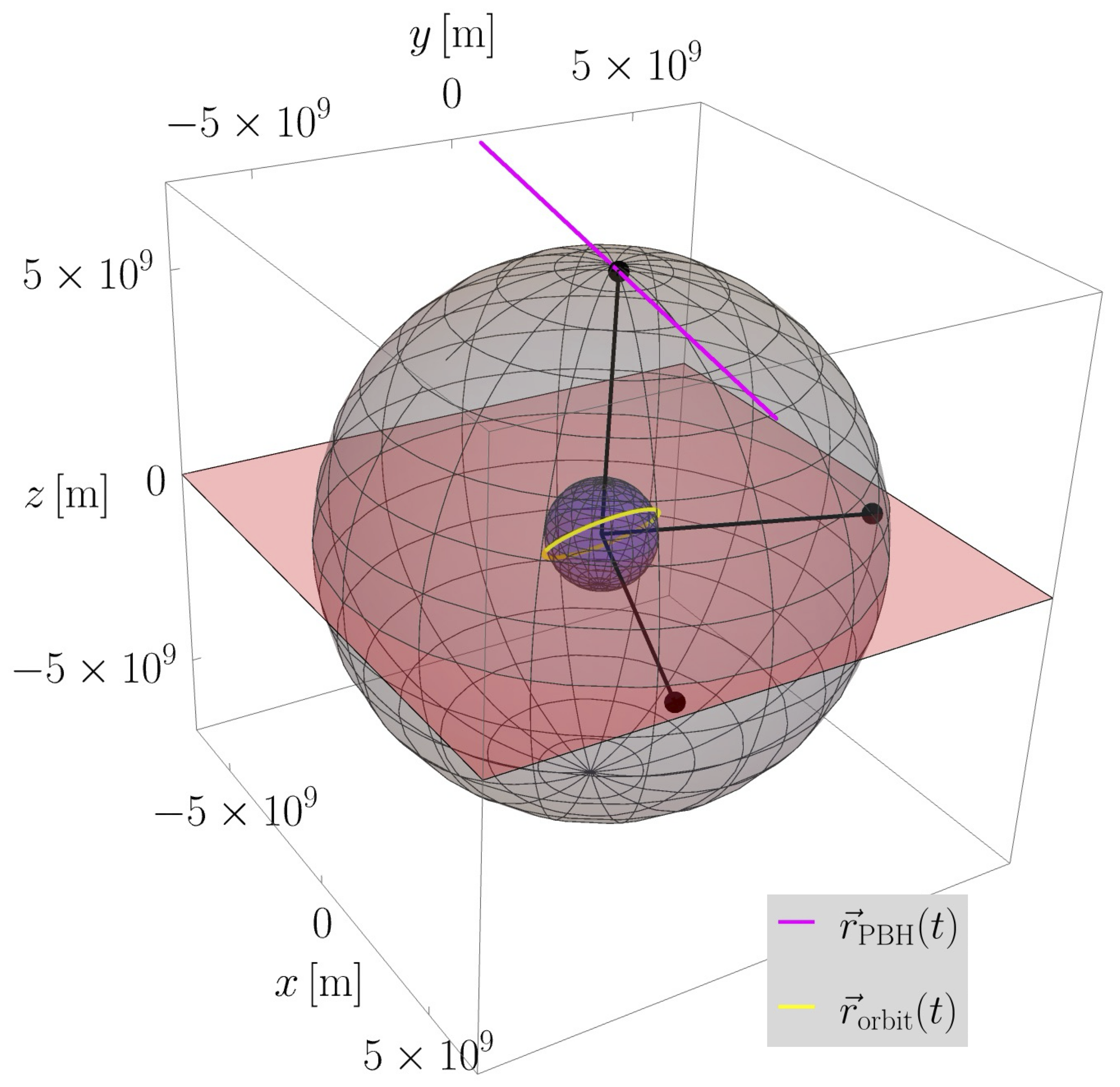}
    \caption{\justifying Example of a simulation geometry for a PBH trajectory with impact parameter $b$ (magenta line) and a detector in low-Earth orbit (yellow ellipse) about the Earth (blue sphere) centered at the origin. We take the the orbital inclination to be with $\theta_{\rm inc}=45^\circ$. The equatorial plane is shown in red and the gray sphere with radius $b$ is to guide the eye. The sizes of the Earth and satellite orbit have been scaled up to be visible. See Fig.~\ref{fig:Zeta1} for the measured photon count rate signal from this example geometry with arbitrary normalization.}
    \label{fig:ExGeometry}
\end{figure}

\begin{figure}[t]
    \centering
    \includegraphics[width=0.49\textwidth]{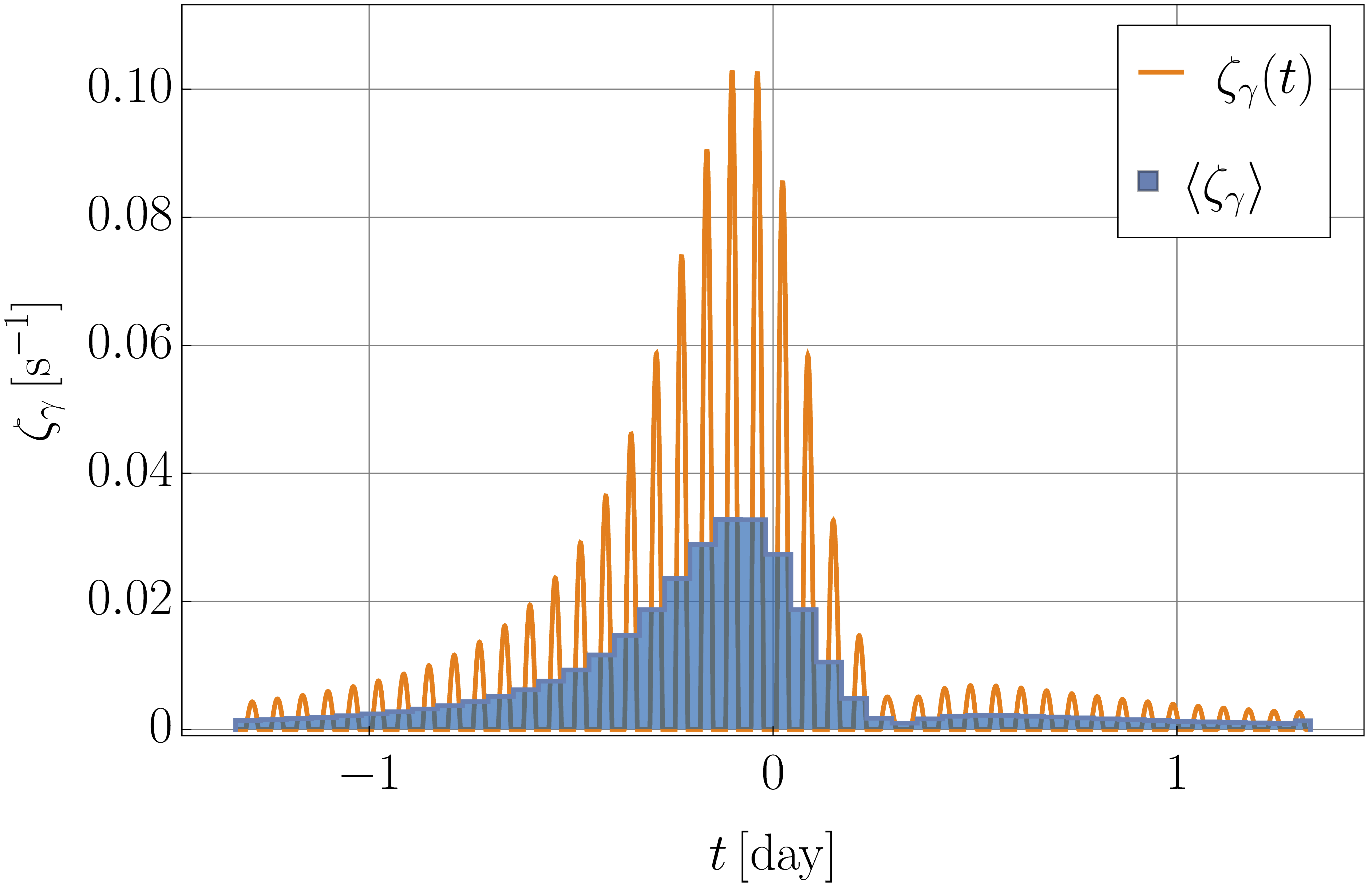}
    \caption{\justifying Measured photon count rate given by Eq.~(\ref{MeasPhotonCountRate}) for the simulation geometry shown in Fig.~\ref{fig:ExGeometry} (orange). The average photon count rate, binned over intervals of one detector orbital period $\tau$, is shown with the blue histogram. Note that a very small value of $b=10^3 R_\oplus= 6.4\times 10^9 \, {\rm m}\sim 0.4 \, {\rm AU}$ is used to easily resolve the oscillatory signal for the sake of this example. We take the detector FOV to be $\theta_{\rm FOV}=\pi/2$ and arbitrarily set $dN_\gamma/dt=10^{20}\,{\rm s}^{-1}$. The PBH velocity for this transit is $v=271.5 \, {\rm km/s}$. Asymmetrical signals like this are possible when the path of the PBH is not in the plane of the detector orbit.}
    \label{fig:Zeta1}
\end{figure}

The parametric PBH trajectory can thus be defined by
\begin{equation}
    \label{PBHParamPath}
    \vec{r}_{\rm PBH}(t|b, \theta_1, \phi_1, \psi_2, v) = \vec{r}_ 1 - \
\frac{\vec{r}_ 2}{|\vec{r}_ 2|}vt,
\end{equation}
with $v$ drawn from the distribution $f(v)$ in Eq.~(\ref{eqn:Maxwellian}). The path is parameterized such that $t=0$ corresponds to the point of closest approach to the origin.

We consider a detector in circular orbit about the Earth with sensitive energy band $[E_{\rm min}, \, E_{\rm max}]$, orbital height $h$, orbital inclination $\theta_{\rm inc}$, orbital period $\tau$, energy-dependent detection efficiency $\epsilon(E)$, and radially outward oriented geometric area $\vec{A}_{\rm geo}$. The orbit is parameterized in Cartesian coordinates by
\begin{equation}
    \label{DetParamPath}
    \begin{split}
    \vec{r}_{\rm orb}(t) & = (h + R_{\oplus})\bigg\langle \cos\left(\frac{2 \pi}{\tau}t+\psi\right)\cos(\theta_{\rm inc}), \\
    & \, \sin\left(\frac{2 \pi}{\tau}t+\psi\right), \, \cos\left(\frac{2 \pi}{\tau}t+\psi\right)\sin(\theta_{\rm inc}) \bigg\rangle,
    \end{split}
\end{equation}
where $\psi\in [0, 2\pi)$ is a randomly sampled phase.
We are working in Cartesian coordinates centered on the Earth with the equator lying in the $x-y$ plane and $\hat{z}$ pointing north. Figure \ref{fig:ExGeometry} shows an example transit geometry (with the Earth radius not to scale), including a PBH trajectory, detector orbit, and the Earth.

The time-dependent photon count rate measured by the detector is given by 
\begin{equation}
    \zeta_{\gamma}(t) = A_{\rm geo} \left(\frac{dN_{\gamma}}{dt}\right)_{\rm det}\frac{\vec{r}_{\rm orb}(t)}{|\vec{r}_{\rm orb}(t)|} \cdot \frac{\vec{\scriptr}(t)}{4\pi \vec{\scriptr}^3(t)} \Theta(\theta_{\rm FOV} - \theta(t)),
    \label{MeasPhotonCountRate}
\end{equation}
where
\begin{equation}
    \vec{\scriptr}(t) \equiv \vec{r}_{\rm PBH}(t|\vec{r}_ 1, \vec{r}_ 2, v) - \vec{r}_{\rm orb}(t|h, \tau, \theta_{\rm inc})
\end{equation}
and $\theta(t)$ is the angle between the two vectors:
\begin{equation}
    \theta(t) \equiv  \cos^{-1}\left( \frac{\vec{\scriptr}(t) \cdot \vec{r}_{\rm orb}(t)}{|\vec{\scriptr}(t)||\vec{r}_{\rm orb}(t)|}\right).
\end{equation}
The emission rate for detectable photons, $(dN/dt)_{\rm det}$, is given by Eq.~(\ref{DethawkPhotons}). Note that the Heaviside theta function $\Theta$ imposes the condition that there is no signal when the PBH is outside the field-of-view (FOV) cone of the detector. The time-dependent measured photon count rate $\zeta_{\gamma}(t)$ given by Eq.~(\ref{MeasPhotonCountRate}) depends on 11 additional parameters besides time, which are listed in Table \ref{tab:ZetaParams}.

\begin{table*}
    \begin{ruledtabular}
    \begin{tabular}{c  c  c  c  c  c  c }
    \>band\> & $\Delta\lambda \, {\rm [\AA]}$ & \>$\Delta E_\gamma \, {\rm [eV]}$\> & \>$\theta_{\rm FOV}$\>  & $S_\nu\, {\rm [Jy]}$ & \> $M \, {\rm [g]} \>$ &\>$(dN_\gamma / dt)_{\rm det} \, {\rm [ s^{-1}]} \>$\\
    \hline
    NUV & 1771-2831 & $4.4 - 7.0 $ & $0.64^\circ$ &  $1.8\times10^{-8}$&\>$\{ 9.1 - 14.6\} \times 10^{21}$\> & \> $\{1.6 - 3.0  \} \times 10^{12}$ \>\\
    FUV & 1344-1786 & $6.9 - 9.2$ & $0.62^\circ$ & $1.6\times10^{-9}$ &  \>$\{6.9 - 9.2\} \times 10^{21}$\> & \> $\{ 2.0 - 2.9 \} \times 10^{12}$ \>\\
    \end{tabular}
    \end{ruledtabular}
    \vspace{5pt}
    \caption{\justifying Some characteristics of the Near Ultraviolet (NUV) and Far Ultraviolet (FUV) detectors on the GALEX satellite~\cite{Bianchi:2013pxa}. 
    The sensitivities $S_\nu$ are taken from Ref.~\cite{sobrinhoDirectDetectionBlack2014}. The sixth column reports the range of PBH masses for which the primary Hawking photon spectrum peaks in the detector's energy band $\Delta E_\gamma$. The last column reports the corresponding emission rate of detectable photons for PBHs in the given mass ranges.
    }
    \label{tab:GALEXinfo}
\end{table*}

\begin{figure}[t]
    \centering
    \includegraphics[width=0.49\textwidth]{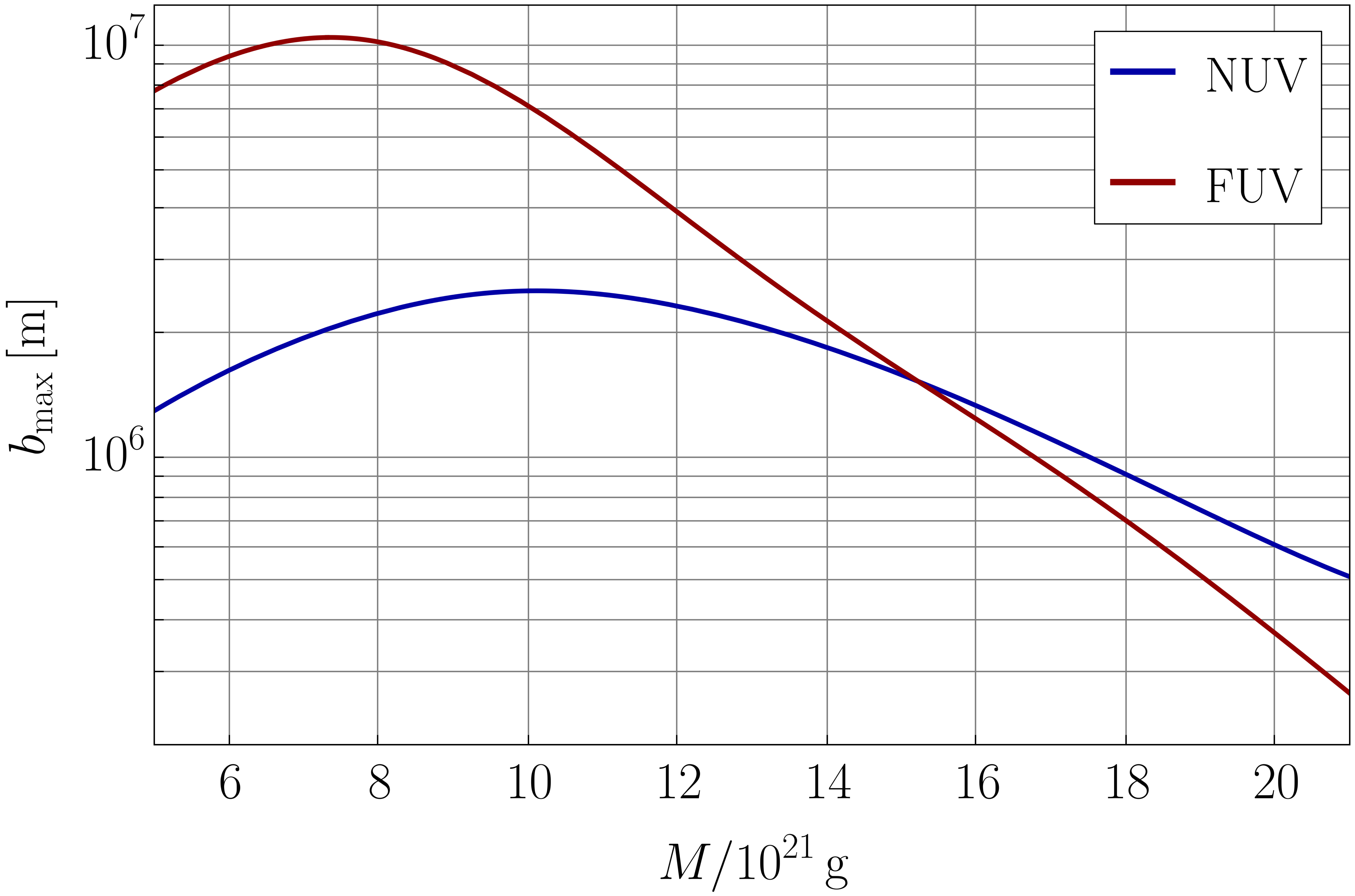}
    \caption{\justifying Maximum detectable impact parameter $b_{\rm max}$ as a function of PBH mass for a PBH \emph{stationary} relative to the GALEX detector. The PBH masses that maximize the curves are $M_{\rm max}^{\rm NUV}=1.01\times10^{22}\,{\rm g}$ ($b_{\rm max}^{\rm NUV}=2.5\times10^6\,{\rm m}$) and $M_{\rm max}^{\rm FUV}=7.34\times10^{21}\,{\rm g}$ ($b_{\rm max}^{\rm FUV}=1.1\times10^7\,{\rm m}$).}
    \label{fig:GALEXbmaxM}
\end{figure}

See Fig.~\ref{fig:Zeta1} for a plot of the measured count-rate $\zeta_{\gamma}(t)$ for the geometry shown in Fig.~\ref{fig:ExGeometry}. Note that for this plot, the count normalization is arbitrary and parameter values have been chosen to make the oscillations visible. Unlike Ref.~\cite{Klipfel:2025bvh}, which only considered trajectories in the plane of the detector orbit, we can now observe highly asymmetrical count rate signals depending on how the PBH trajectory is oriented relative to the detector orbital plane. Figure~\ref{fig:Zeta1} also overlays a histogram of the average count rate binned over intervals of one orbital period $\tau$ with bin heights:
\begin{equation}
    \langle \zeta_\gamma\rangle_i=\frac{1}{\tau}\int_{t_i}^{t_i+\tau}dt \,\zeta_\gamma(t),
\end{equation}
where $t_i$ are the lower bin edges.

\subsection{Anticipated Signals for Space-based Experiments}
\label{sec:SpaceBased}
We consider detection of Hawking photons by three different instruments: the near-UV and far-UV detectors on the GALEX satellite (GALEX-NUV and GALEX-FUV), as well as AMEGO-X.

\subsubsection{GALEX}

The {\it Galaxy Evolution Explorer} (GALEX) satellite mission was launched in April 2003, with instruments to observe in the near-UV (NUV, $1771 - 2831$ \AA\>) and far-UV (FUV, $1344 - 1786$ \AA \>) \cite{Bianchi:2013pxa}. The FUV detector malfunctioned in 2009 and the mission continued observing in the NUV until 2013. See Ref.~\cite{Bianchi:2013pxa} for a summary of All-sky and Medium-depth Imaging Surveys, Ref.~\cite{Gezari:2013wkj} for a detailed discussion of the GALEX Time Domain Survey, and Ref.~\cite{Morrissey:2007hv} for a discussion of GALEX calibration and detector parameters. Table~\ref{tab:GALEXinfo} includes GALEX parameters referenced in this section for the NUV and FUV instruments.

A first estimate of the prospects for detecting PBHs with GALEX was performed by Ref.~\cite{sobrinhoDirectDetectionBlack2014}, which derived a maximum detectable separation distance between a stationary PBH and the detector of $\mathcal{O}(10^6/10^7\,{\rm m})$ (NUV/FUV). However, as we note in Ref.~\cite{Klipfel:2025bvh}, one expects PBHs to be gravitationally bound to the galaxy rather than the Solar System and thus to have a large velocity relative to the Earth. In this section, we evaluate the prospects for detecting a \emph{moving} PBH with existing GALEX data. 

The narrow GALEX field-of-view diameter of $1.28^\circ/1.24^\circ$ (NUV/FUV), high angular resolution of $5.3''/4.2''$ (NUV/FUV), and long exposure times of $\sim150-1500 \, {\rm s}$ are well-suited for the localization of distant sources such as stars and galaxies \cite{Bianchi:2013pxa}. However, as we will show, these properties make GALEX a non-ideal candidate experiment for PBH transit detection when paired with the modest Hawking emission rates by all PBHs in the NUV and FUV and the large relative velocities between the PBHs and the instrument. 

We first estimate the maximum detectable distance for a stationary PBH in the NUV and FUV. We then compute the proper motion of such a source given a typical PBH trajectory through the inner Solar System and evaluate whether GALEX would be able to resolve the moving PBH as a point source in a single exposure, and whether it could register as a transient in the GALEX Time Domain Survey (TDS) \cite{Gezari:2013wkj}. 

Following Ref.~\cite{sobrinhoDirectDetectionBlack2014}, the flux density $S_{\nu}$(M) in Janskys is related to the PBH Hawking emission spectrum and the impact parameter $b$ by
\begin{equation}
\begin{split}
    S_\nu & =\frac{\pi r_s^2}{b^2}\frac{1}{\Delta\nu}\int_{\nu_{\rm min}}^{\rm \nu_{\rm max}}d\nu \, B_\nu(M) \\
    & = \frac{1}{4b^2}\frac{2\pi \hbar}{\Delta E}\int_{E_{\rm min}}^{E_{\rm max}}dE\,E \frac{d^2N_{\gamma}^{(2)}}{dtdE},
    \end{split}
    \label{Snu}
\end{equation}
where $r_s = 2 G M$ is the PBH Schwarzschild radius.
The typical detector sensitivity in a band $i=$ FUV, NUV is the flux density $S_\nu^i$ corresponding to the minimum resolvable magnitude of a point source at the central frequency of the band. Taking the values from Ref.~\cite{sobrinhoDirectDetectionBlack2014}, we have $S_\nu^{\rm NUV}=1.8\times10^{-8}\,{\rm Jy}$ and $S_\nu^{\rm FUV}=1.6\times10^{-9}\,{\rm Jy}$. Setting $S_\nu=S_\nu^i$ in Eq.~(\ref{Snu}) and solving for $b_{\rm max}$ gives the maximum detectable impact parameter in a given band as a function of PBH mass $M$:
\begin{equation}
    b_{\rm max}^i(M) = \left(\frac{1}{4S_\nu^i}\frac{2\pi \hbar}{\Delta E}\int_{E_{\rm min}}^{E_{\rm max}}dE\,E \frac{d^2N_{\gamma}^{(2)}}{dtdE}\right)^{1/2}.
\end{equation}
Figure~\ref{fig:GALEXbmaxM} plots $b_{\rm max}^i(M)$ for the NUV and FUV bands. We find that the PBH masses that maximize the curves are $M_{\rm max}^{\rm NUV}=1.01\times10^{22}\,{\rm g}$ ($b_{\rm max}^{\rm NUV}=2.5\times10^6\,{\rm m}$) and $M_{\rm max}^{\rm FUV}=7.34\times10^{21}\,{\rm g}$ ($b_{\rm max}^{\rm FUV}=1.1\times10^7\,{\rm m}$). Note that the typical Earth-moon distance is $3.8\times10^8 \,{\rm m}$.

We now account for the proper motion of a PBH on a typical trajectory with velocity $v\sim250 \, {\rm km/s}$. The GALEX TDS reports typical exposure times between $\sim150-1500\,{\rm s}$ with minimum and maximum exposure lengths of $\sim 30 \, {\rm s}$ and $\sim 2.2\times10^4 \, {\rm s}$. Taking the shortest possible exposure times of $t_{\rm exp}=32\,{\rm s}/31\,{\rm s}$ (NUV/FUV), and the furthest possible distance for a detectable signal $b_{\rm max}^i$, we can estimate a lower bound on the proper motion (in degrees) of a PBH on a trajectory parallel to the plane of the detector:
\begin{equation}
    \theta_{\rm prop}^i = 2\tan^{-1}\left(\frac{v \, t_{\rm exp}}{2 b_{\max}^i} \right).
\end{equation}

The minimum attainable values for the proper motion are thus $\theta_{\rm prop}^{\rm NUV}=57.5^\circ$ and $\theta_{\rm prop}^{\rm FUV}=20.3^\circ$, which are exponentially larger than the angular resolution of $\sim10^{-3}$ degrees. Thus, because $b_{\rm max}^i$ is so small for either band, the PBH will traverse a large fraction of the sky during the duration of a single GALEX exposure and will not be detected as a clean point source in even a single frame for a transient analysis such as the TDS. For reference, the distance such that the PBH would appear to have proper motion smaller than the angular resolution is $b\sim 10^{11}\, {\rm m}\simeq 1 \, {\rm AU}$---but the photon signal would be negligible at such a distance. 

Due to this large proper motion for $b\leq b_{\rm max}$, the only way GALEX could detect a PBH is if the trajectory were along the detector line of sight, which would result in the PBH impacting the Earth (or the instrument itself). For a PBH number distribution that is sharply peaked at $\bar{M} = 10^{22}\,{\rm g}$ and a local dark matter density of $0.0155 \, M_\odot {\rm pc}^{-3}$, the Earth impact rate is $\sim 10^{-13}\,{\rm yr}^{-1}$.  
We therefore conclude that PBH transit detection with GALEX is not feasible. Given such exponentially low event rates, we do not perform transit simulations for detections by GALEX.

\subsubsection{AMEGO-X}

An instrument of interest to detect higher-energy photons from PBH transits is the proposed {\it All-sky Medium Energy Gamma-ray Observatory eXplorer} (AMEGO-X) satellite. According to Ref.~\cite{Martinez-Castellanos:2021bbl},
the satellite's instruments should be able to track photons from transient phenomena within the range $10 \, {\rm keV} \leq E_\gamma \leq 1 \, {\rm GeV}$ with typical sensitivities of $\sim 0.5-11 \,{\rm cm}^{-2}{\rm s}^{-1}$ for a $1 \, {\rm s}$ duration (see Fig.~9 from Ref.~\cite{Martinez-Castellanos:2021bbl}) and localize them to within $\sim1^\circ$ on the sky. See Table~\ref{tab:AXDetInfo} for a list of relevant parameters for the AMEGO-X instrument and its orbit. 

\begin{figure}[t]
    \centering
    \includegraphics[width=0.49\textwidth]{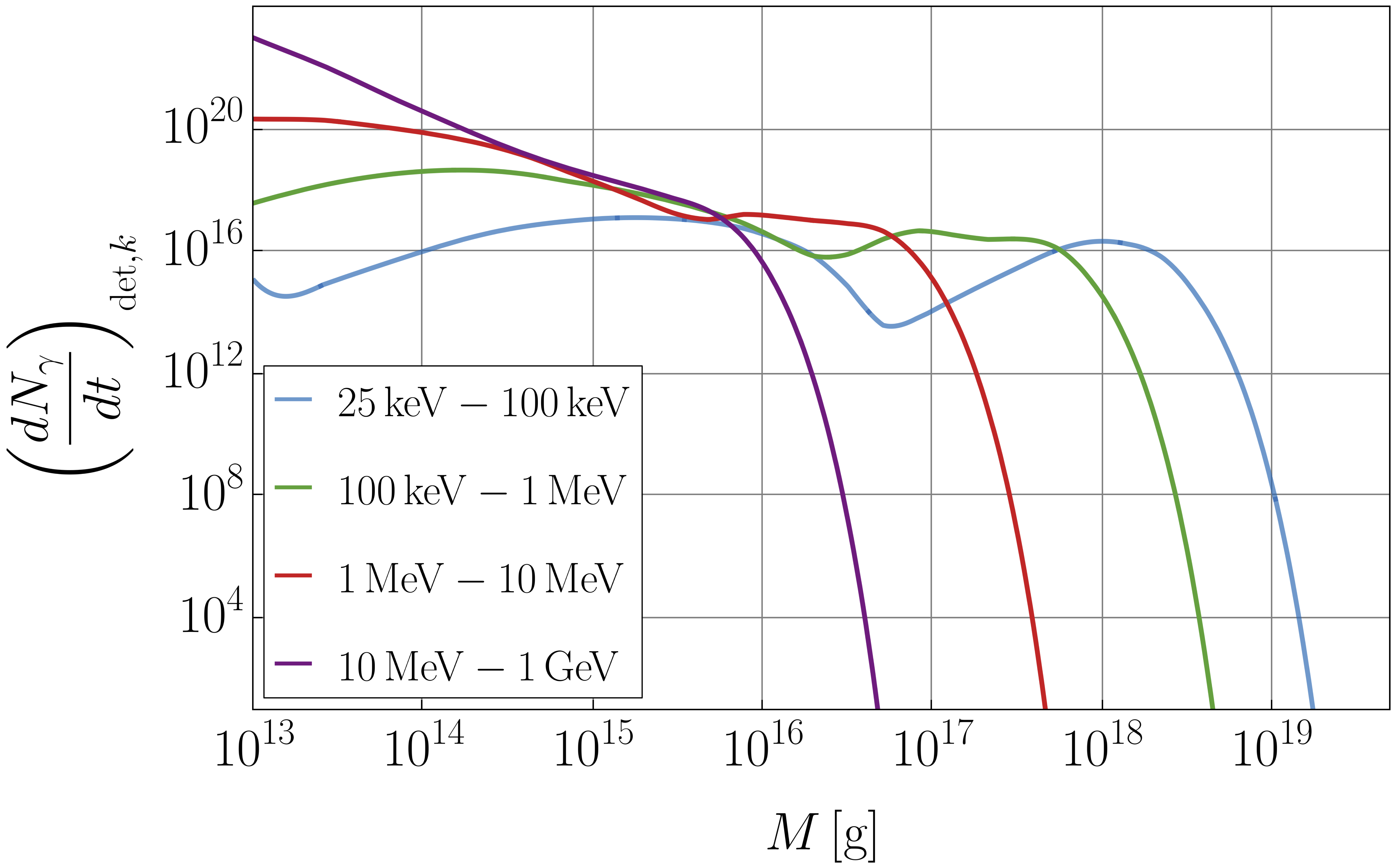}
    \caption{\justifying Emission rate of \emph{detectable} photons in the four energy bins $\Delta E_k$ expected for the proposed AMEGO-X experiment computed by integrating the secondary spectra shown in Fig.~\ref{fig:HawkPrimary} according to Eq.~(\ref{DethawkPhotons}), with the detector efficiency derived from Ref.~\cite{Martinez-Castellanos:2021bbl}.}
    \label{fig:AXdetrate}
\end{figure}

\begin{table*}
    \begin{ruledtabular}
    \begin{tabular}{c  c  c  c  c  c  c }
    $\Delta E$ & $A_{\rm geo} \, {\rm [cm^2]}$ & $h \, {\rm [km]}$ & $\theta_{\rm inc} $ & $\theta_{\rm FOV} \, {\rm [rad]}$ & $\tau \, {\rm [min]}$ & $M \, {\rm [g]}$ \\
    \hline
    $25\ {\rm keV} - 1 \, {\rm GeV}$ & $6400$ & $575$ & $6^\circ$ & $\pi/2$ & $95$ & $M\lesssim3\times10^{18}$\\
    \end{tabular}
    \end{ruledtabular}
    \vspace{5pt}
    \caption{\justifying Some characteristics of the proposed AMEGO-X satellite~\cite{Caputo:2022xpx, Martinez-Castellanos:2021bbl} used in our simulations.} 
    \label{tab:AXDetInfo}
\end{table*}

We note that AMEGO-X has the ideal energy range to perform a follow-up analysis to our work in Ref.~\cite{Klipfel:2025bvh}, which simulated time-series positron signals from PBH transits through the inner Solar System with $M\lesssim 5\times10^{17}\,{\rm g}$. AMEGO-X would be sensitive to photons emitted by PBHs near the lower end of the asteroid-mass range, with $M \lesssim 3 \times 10^{18} \, {\rm g}$. (See also Ref.~\cite{Ray:2021mxu}.) Note that throughout the range $2\times10^{17} \, {\rm g} \lesssim M \lesssim 6 \times 10^{18} \, {\rm g}$, there do not presently exist {\it any} constraints on $f_{\rm PBH}$, when considering monochromatic PBH number distributions. 

The AMEGO-X gamma ray telescope (GRT) is capable of detecting four distinct types of photon events, defined by how the photon interacts with the detector. The GRT consists of a $40$-layer $6400\, {\rm cm}^2$ silicon pixel tracker (four modules $40\times40\,{\rm cm}^2$ each), a $4$-layer Cesium Iodide hodoscopic calorimeter, and a plastic scintillator anti-coincidence counter. Single site events (SSE) occur when a low-energy photon is absorbed via the photoelectric effect and deposits all its energy to one pixel of one layer of the silicon pixel tracker. This type of event has the highest background. Tracked and un-tracked Compton events (TC and UC) occur when a medium-energy photon Compton scatters with an electron in some layer of the tracker and the path of the electron can either be reconstructed (tracked) via its interaction with other tracker layers or not (un-tracked). Pair-production events (P) occur for the most energetic $\gamma$-rays, which can emit an $e^{\pm}$ pair after interacting with the material of the tracker. Each event type dominates over a typical energy range. We use this to define four energy bins that roughly correlate with regimes in which different event-types dominate: $\Delta E_{\rm SSE}=25\,{\rm keV}-100\,{\rm keV }, \, \Delta E_{\rm UC}=100\,{\rm keV }-1\,{\rm MeV}, \, \Delta E_{\rm TC} = 1\,{\rm MeV}-10\,{\rm MeV},\, \Delta E_{P}=10\,{\rm MeV}-1\,{\rm GeV}$. See Table~\ref{tab:AXparams}.

For a given PBH transit with mass $M$ and impact parameter $b$, we compute $\left( dN_\gamma/dt\right)_{{\rm det}, k}$ for each energy bin $k$, and then simulate the time-dependent photon signal at the detector separately for each energy bin following the method described in Section~\ref{sec:PhotonSignals}. In the simulations we use detector parameters in Table~\ref{tab:AXDetInfo} and the reported $A_{\rm eff}$ from Ref.~\cite{Martinez-Castellanos:2021bbl}. Bin-averaged values of $\epsilon(E)\equiv A_{\rm eff}(E)/A_{\rm geo}$ are shown in Table~\ref{tab:AXparams}. Incorporating the detection efficiency when computing $\left( dN_\gamma/dt\right)_{{\rm det}, k}$ reduces the typical emission rate in an energy-dependent (and thus PBH mass-dependent) way. The rates are scaled down by a factor between $3-50$ depending on the PBH mass. This will result in a reduced (yet more accurate) signal compared to simulations which do not account for detection efficiency.

Given a randomly sampled velocity, we simulate the signal over an interval of time $[-t_{\rm max}, t_{\rm max}]$, where $t_{\rm max}(b, v)$ satisfies:
\begin{equation}
    \sqrt{b^2 + (v \,t_{\rm max})^2}=\frac{b}{10}.
\end{equation}
This enforces that the PBH signal increases and then decreases by a factor of $10^2$ throughout the duration of the simulation. (Recall that the point of closest approach occurs for $t=0$.) 

\begin{figure}
    \includegraphics[width=0.49\textwidth]{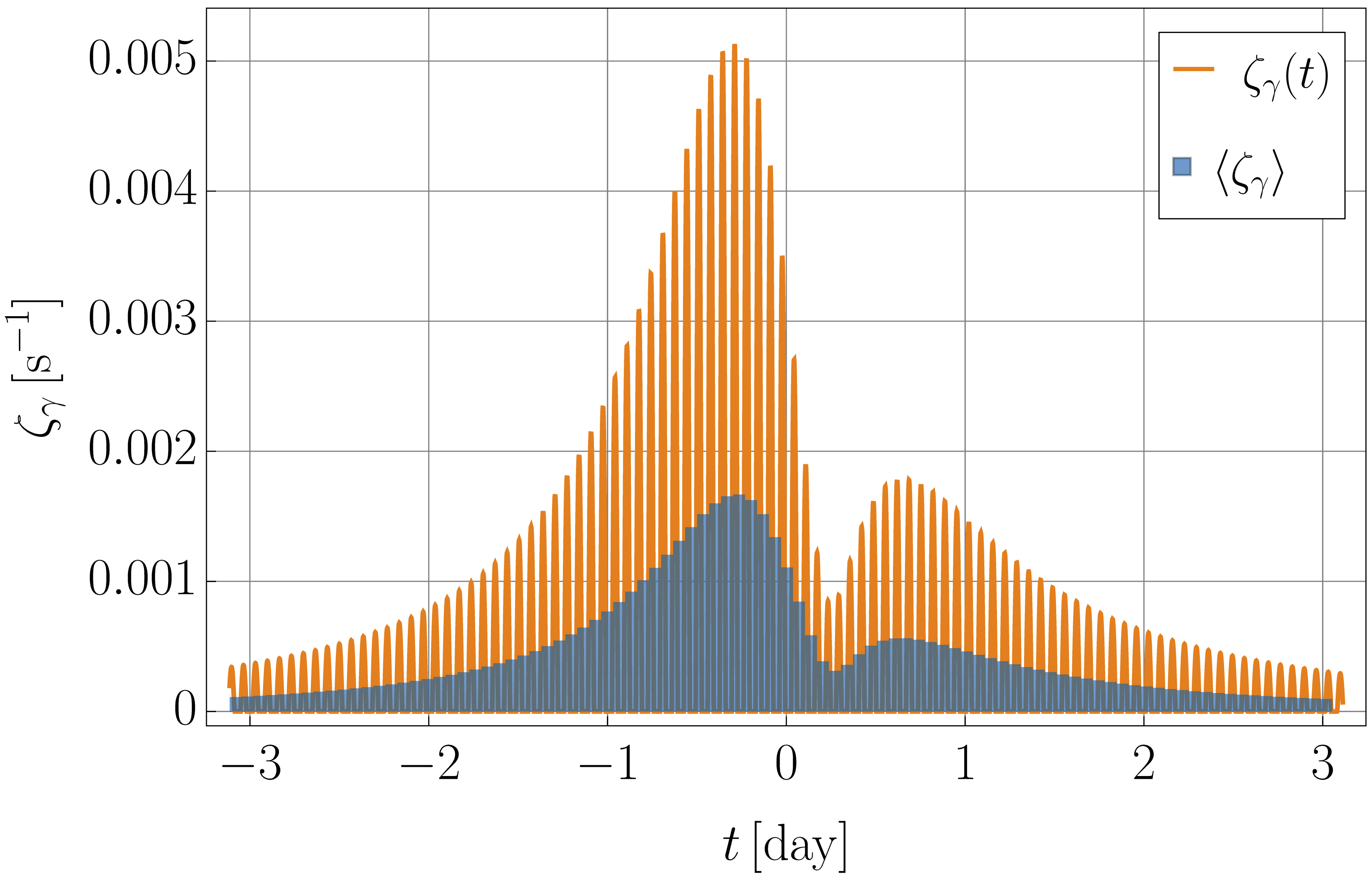} \includegraphics[width=0.49\textwidth]{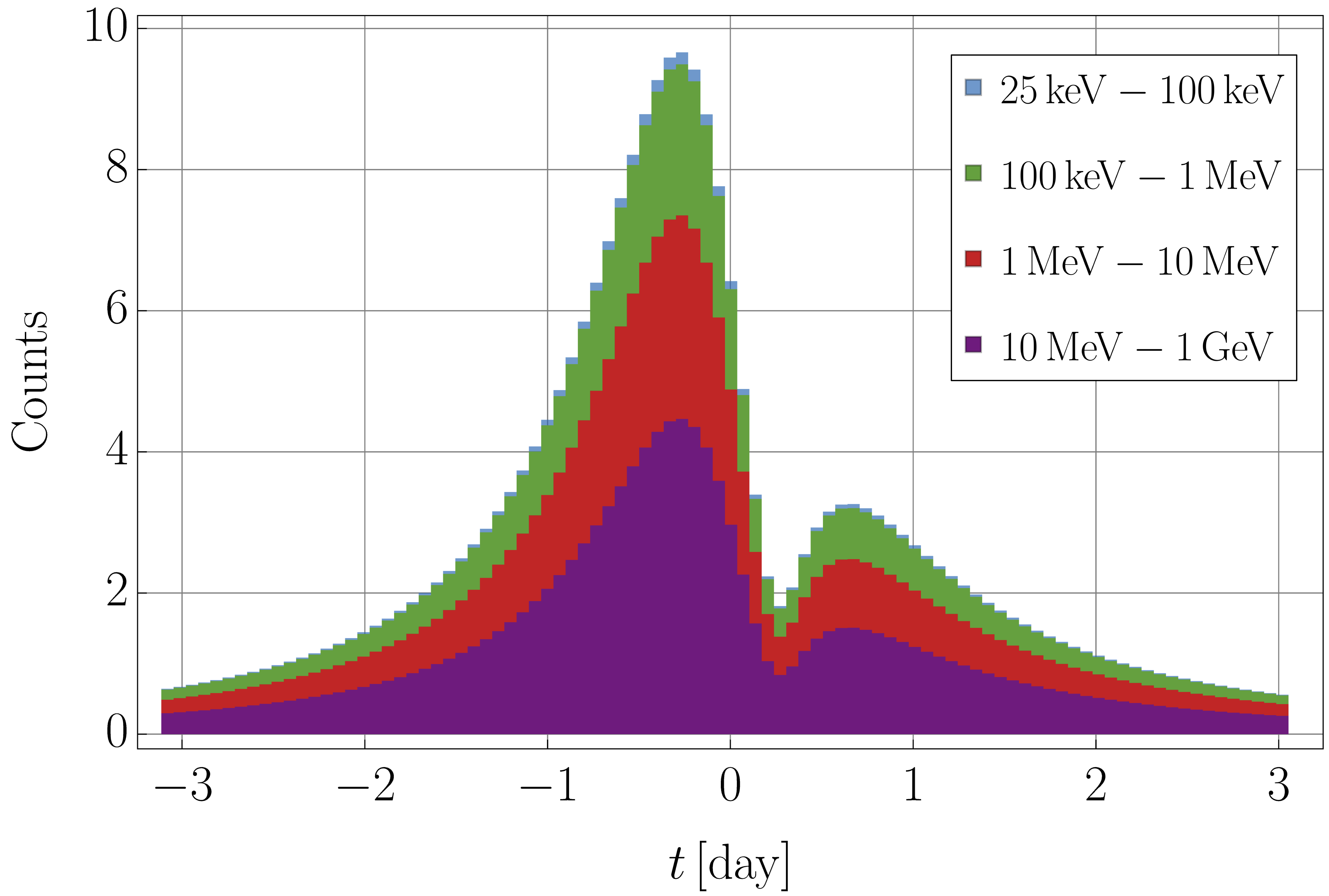}
    \caption{\justifying Simulated transit signal with parameters $M=10^{15}\,{\rm g}$, $b=0.04\, {\rm AU}$ for the AMEGO-X detector, with $v= 116 \,{\rm km/s}$ selected randomly from the distribution in Eq.~(\ref{eqn:Maxwellian}) for the particular transit shown. (\emph{Top}) The total measured photon count rate as a function of time $\zeta_\gamma(t)$ and the average count rate binned on time intervals of one orbit $\tau$. (\emph{Bottom}) A stacked histogram of total counts in each of the four energy bins. Note that the particular shape of the signal here occurs because the PBH trajectory was such that the signal was occluded by the Earth around the time of closest approach, resulting in two peaks. 
    }
    \label{fig:AXzeta}
\end{figure}

This results in a time-series signal,
\begin{equation}
    \boldsymbol{\zeta}(t) =\{\zeta_k(t)\},
\end{equation}
where $k$ runs over all four energy bins. We then integrate the rate over time intervals of one detector orbit $\tau$ to get a time-dependent signal of photon \emph{counts} binned in time with bin heights
\begin{equation}
    Z_k^j = \int_{t_j-\tau/2}^{t_j+\tau/2}dt \, \zeta_k(t),
\end{equation}
where the central value of the $j$th bin is $t_j = -t_{\rm max} + \tau(j-\frac{1}{2})$. See Fig.~\ref{fig:AXzeta}a for a plot of the time-dependent total photon count rate $\sum_k\zeta_k(t)$ and Fig.~\ref{fig:AXzeta}b for a histogram of photon counts in each bin $Z_k^j$ for an example transit signal measured by AMEGO-X.

The signal-to-noise ratio (SNR) for bin $Z_k^j$ is \cite{Martinez-Castellanos:2021bbl}
\begin{equation}
    {\rm SNR}_k^j=\frac{Z_k^j}{\sqrt{Z_k^j + B_k^j}},
\end{equation}
where the expected background count for the bin is 
\begin{equation}
    B_k^j = \tau \int_{E_k^{\rm min}}^{E_k^{\rm max}}dE \, \frac{d^2 \mathcal{B}}{dEdt},
    \label{eq:Bkj}
\end{equation}
and the expected background flux $d^2\mathcal{B}/dEdt$ is taken from Fig.~4 of Ref.~\cite{Martinez-Castellanos:2021bbl}. See Table~\ref{tab:AXparams} for bin-averaged values of the background flux.

The maximum SNR for each energy bin is found by 
\begin{equation}
    {\rm SNR}_k = \max_j{\left\{\frac{Z_k^j}{\sqrt{Z_k^j+B_k^j}}\right\}}.
\end{equation}
Following Ref.~\cite{Martinez-Castellanos:2021bbl}, we can then add the maximum SNR from each energy bin in quadrature to achieve a total SNR for the transit:
\begin{equation}
    {\rm SNR}_{\rm tot} = \left(\sum_k {\rm SNR}_k^2 \right)^{1/2}.
\end{equation}
Thus, detecting a signal in multiple energy bins improves the statistical power of a detection. 

Using some threshold SNR for a transit detection $S_{\rm thresh}\in\{3, 10^{-1}, 10^{-2}, 10^{-3} \}$, we define a detection to be  
\begin{equation}
    \mathcal{D}(S_{\rm thresh})\equiv \begin{cases}
        1 & {\rm if} \quad {\rm SNR}_{\rm tot} \geq S_{\rm thresh}\\
        0 & {\rm otherwise.} \\
        \end{cases}
\end{equation}
Ref.~\cite{Martinez-Castellanos:2021bbl} uses a threshold SNR of 6.5 for detecting gamma-ray bursts, to ensure a false detection rate of $<1 \, {\rm yr}^{-1}$. However, we can consider values of $S_{\rm thresh} < 1$ since application of a matched filter to the full time-series signal $\zeta_k(t)$ can significantly lower the threshold SNR for a detection. As an example, the LIGO-Virgo-KAGRA Collaboration achieves an SNR threshold of $\mathcal{O}(10^{-4})$ with its sophisticated matched filtering techniques and template fits \cite{LIGOScientific:2019hgc}. We therefore report results for three possible levels of $S_{\rm thresh}<1$, corresponding to possible sensitivities with matched filtering within known capabilities. 

\begin{table}[t]
\caption{\label{tab:AXparams} \justifying AMEGO-X energy bins and relevant parameters for each bin, including energy range, bin-averaged detection efficiency $\langle\epsilon\rangle_k\equiv\Delta E_k^{-1}\int dEA_{\rm eff}(E)/A_{\rm geo}$, and bin-averaged background count rate $B_k$. The energy-dependent detection efficiency and background rates are taken from Ref.~\cite{Martinez-Castellanos:2021bbl}.
}
\begin{ruledtabular}
\begin{tabular}{ccc}
$\Delta E_k$ & $\langle\epsilon\rangle_k$   & $B_k \, {[{\rm s}^{-1}]}$ \\ 
\hline
$25\,{\rm keV}-100\,{\rm keV }$ & $0.36$ & $2.19\times10^4$\\
$100\,{\rm keV }-1\,{\rm MeV}$ & $8.9\times10^{-2}$& $4.78\times10^2$ \\
$1\,{\rm MeV}-10\,{\rm MeV}$ &$3.1\times10^{-2}$ & $86.0$\\
$10\,{\rm MeV}-1\,{\rm GeV}$ & $5.9\times10^{-2}$ & $8.55$\\
\end{tabular}
\end{ruledtabular}
\end{table}

\begin{figure*}[t]
    \centering
    \includegraphics[width=0.6\textwidth]{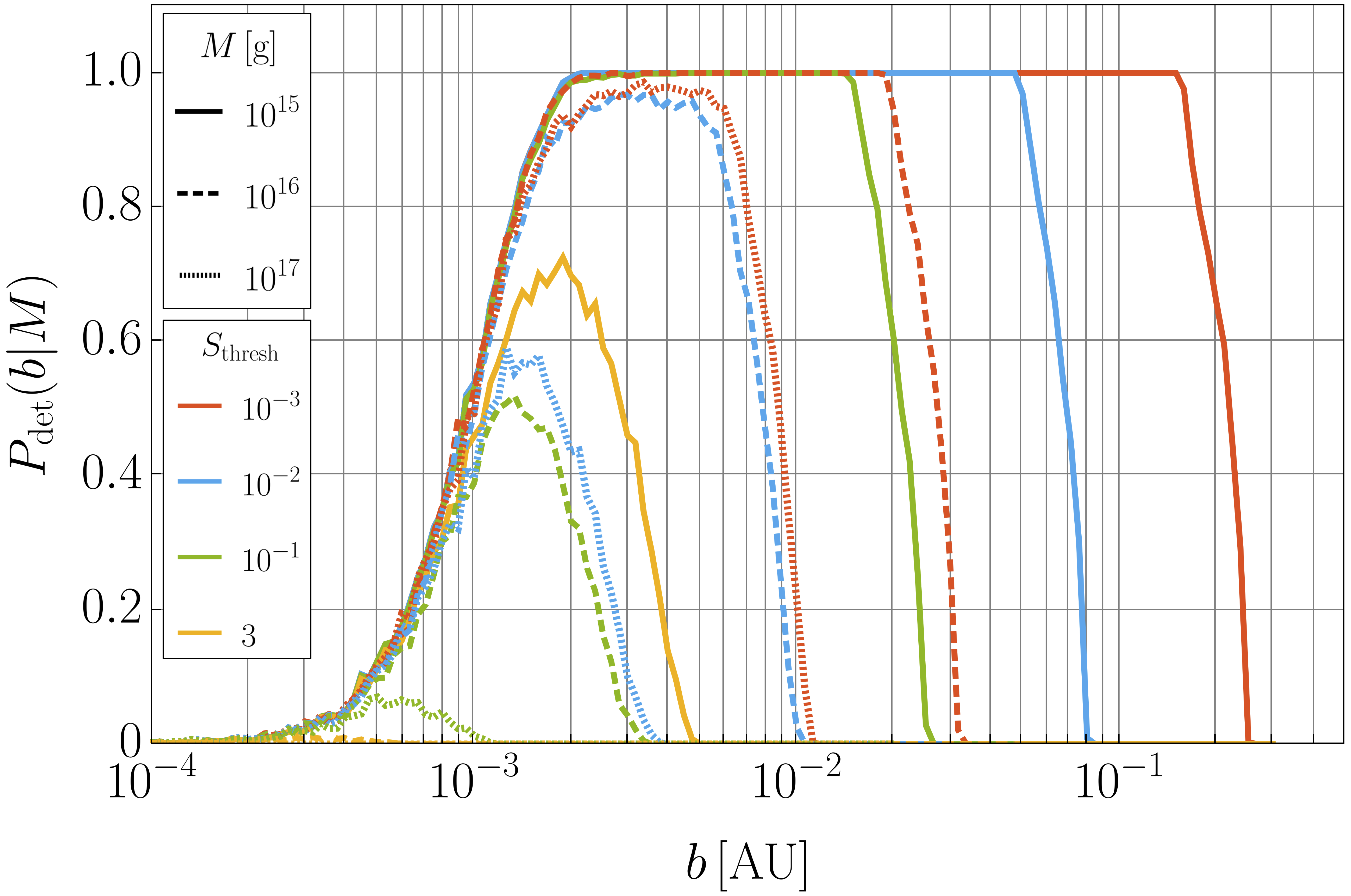}
    \caption{\justifying Probability of PBH transit detection $\mathcal{P}_{\rm det}(b|M)$ for AMEGO-X as a function of impact parameter $b$ and PBH mass $M$, from Eq.~(\ref{eq:PdetMb}). We assume four values of the SNR threshold sensitivity $S_{\rm thresh}$, assuming that detection of transit events with $S_{\rm thresh}<1$ will be possible with matched filtering, as demonstrated in our previous work \cite{Klipfel:2025bvh}. Each point is computed with $n=10^3$ simulations.  The maximum impact parameter for a reliable detection, $b_{\rm max}(M)$ satisfies $\mathcal{P}_{\rm det}(b_{\rm max}|M)=0.99$.}
    \label{fig:b15}
\end{figure*}

For a given PBH mass $M$, we can compute the maximum detectable impact parameter $b_{\rm max}(M)$ by the following procedure. For a parameter set $\{M,b\}$, we run $n$ simulations (where the transit trajectories, orbit phases, and velocities are randomly sampled as discussed in Sec.~\ref{sec:PhotonSignals}),
and compute the detection probability via
\begin{equation}
    \label{eq:PdetMb}
    \mathcal{P}_{\rm det}(M,b|S_{\rm thresh}) = \frac{1}{n}\sum_i^n \mathcal{D}_i(S_{\rm thresh}).
\end{equation}
In Fig.~\ref{fig:b15}, we plot ${\cal P}_{\rm det} (b \vert M)$ for three different PBH masses and four different values of $S_{\rm thresh}$.

The maximum impact parameter $b_{\rm max}$ is defined as the largest value of $b$ such that
\begin{equation}
    \mathcal{P}_{\rm det}(M, b|S_{\rm thresh})=0.99,
\end{equation}
which corresponds to AMEGO-X successfully detecting a PBH of mass $M$, velocity $v\in f(v)$, and $b=b_{\rm max}$ as a transient 99\% of the time. Looking at Fig.~\ref{fig:b15}, we note that there is also a \emph{minimum} impact parameter for a successful detection $b_{\rm min}$, corresponding to the smallest value of $b$ such that $\mathcal{P}_{\rm det}(M, b|S_{\rm thresh})=0.99$. The transit detection efficiency eventually falls for small impact parameter $b$ because the time interval when the PBH is within the field of view of the instrument gets too short for a significant number of photons to hit the detector, despite the higher signal flux at Earth due to the small impact parameter. This trade-off between the $1/b^2$ flux scaling and the transit time interval gives rise to a bounded region $b_{\min}\leq b \leq b_{\rm max}$ within which reliable PBH detection is possible with AMEGO-X given some sensitivity threshold $S_{\rm thresh}$.

\subsection{Detecting Radio-Band Signals on Earth}
\label{sec:Radio}

\begin{table}[t!]
\caption{\label{tab:RadioVals} \justifying Parameters that define endpoints of the radio band we consider for possible PBH Hawking emission signals. The last column lists the PBH mass for which Hawking emission peaks at the given frequency.
}
\begin{ruledtabular}
\begin{tabular}{cccc}
$\lambda$ & $f$ & $E \, [{\rm eV}]$   &  $M \, [{\rm g}]$\\ 
\hline
$0.3\,{\rm mm}$ & $1 \, {\rm THz}$ & $6.6\times10^{-4}$ & $9.7\times10^{25}$ \\
$30\,{\rm m}$ & $10 \, {\rm MHz}$ & $6.6\times10^{-9}$ & $9.7\times10^{30}$ \\
\end{tabular}
\end{ruledtabular}
\end{table}

\begin{figure}[t]
    \includegraphics[width=0.49\textwidth]{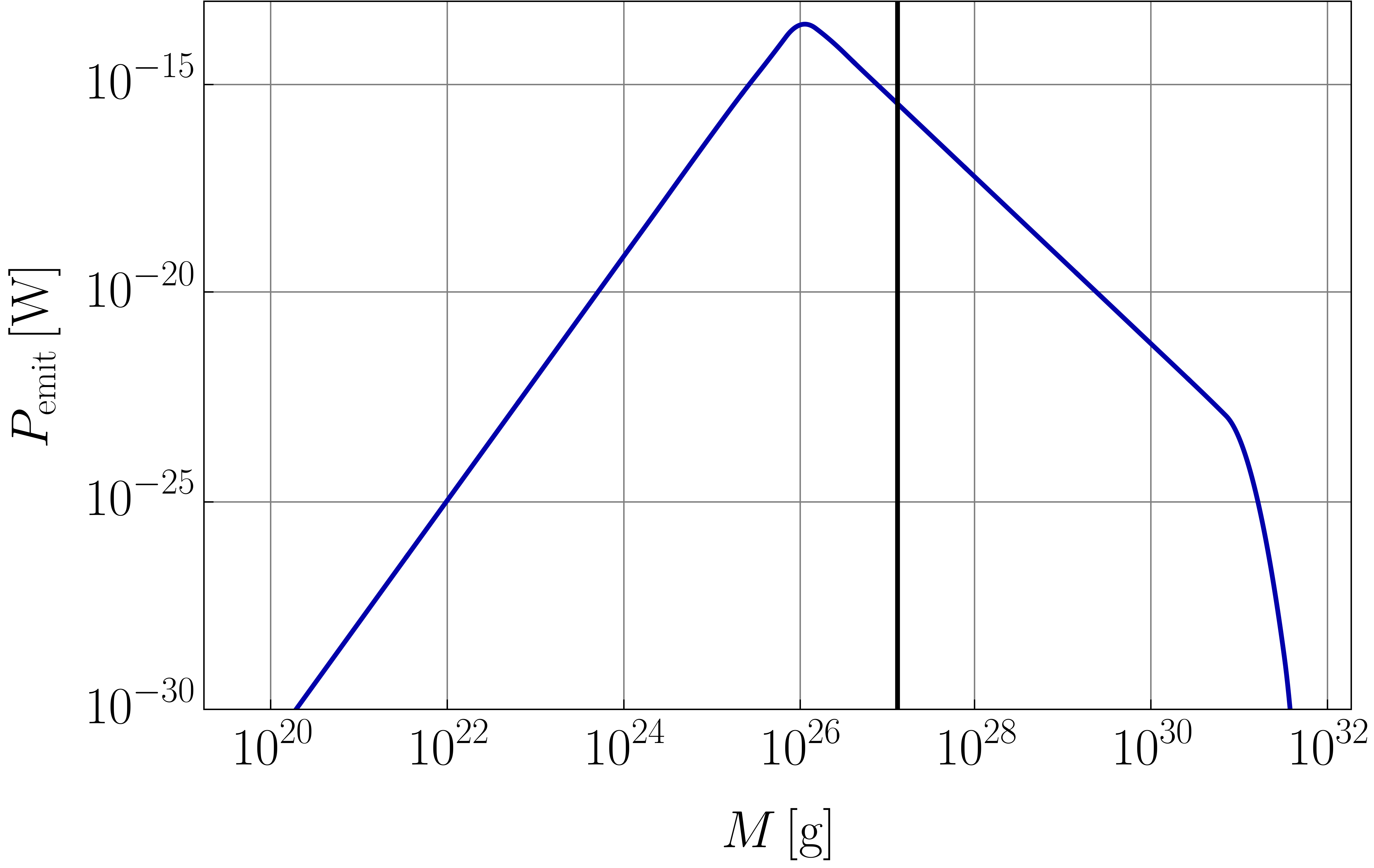} 
    \caption{\justifying Emitted power by a PBH of mass $M$ in the radio band $10 \, {\rm MHz} \leq f \leq 1 \, {\rm THz}$ described in Table~\ref{tab:RadioVals}. The vertical black line corresponds to the critical mass $M_c=1.26\times10^{27}$, such that all PBHs with mass $M\leq M_c$ are net emitters. Note that the upper edge of the asteroid mass window is $M\sim10^{23}\,{\rm g}$.}
    \label{fig:RadioPlot}
\end{figure}

The Earth's atmosphere is transparent to radio waves in the window $0.3 \, {\rm mm} \lesssim\lambda \lesssim 30 \, {\rm m}$ \cite{2016era..book.....C}. We estimate Hawking radiation signals in this band for PBHs with masses at the upper end of the asteroid-mass range and beyond. See Table \ref{tab:RadioVals} for a list of parameters that bound this band.

Following Ref.~\cite{Klipfel:2026aug}, we note that PBHs in the present epoch are not emitting in a vacuum, but rather are immersed in a thermal bath of photons at background temperature $T_b\gtrsim T_{\rm CMB}=2.35\times10^{-4}\, {\rm eV}$. 
Using the criterion for a PBH of mass $M$ to be a net emitter~\cite{Klipfel:2026aug},
\begin{equation}
    M\leq M_c(T_b)\equiv\frac{\pi}{2.82}\frac{1}{GT_b},
\end{equation}
we find that the critical mass for emission if $T_b=T_{\rm CMB}$ is $M_c(T_{\rm CMB})= 1.26\times10^{27}\,{\rm g}$. (For similar considerations on the criteria for net emission, see, e.g., Refs.~\cite{Rice:2017avg,Barrau:2022bfg,Loeb:2024gga}; cf. Ref.~\cite{Chatterjee:2025wnt}.)

For PBHs with mass $M\leq M_c(T_{\rm CMB})$, we can compute the emitted power in the radio window of Table~\ref{tab:RadioVals} via Eq.~\eqref{DethawkPhotons}. We assume $\epsilon=1$ for 100\% efficiency across the band and take $E_{\rm min}$ and $E_{\rm max}$ from column 3 of Table~\ref{tab:RadioVals}. Figure~\ref{fig:RadioPlot} plots the total emitted power across the radio window as a function of PBH mass. Emitted power in the radio band peaks at $P_{\rm emit}=2.91\times10^{-14}\,{\rm W}$ for PBH mass $M=1.04\times10^{26}\,{\rm g}$.

These signals are exponentially too weak to detect. For example, assuming a $70 \, {\rm m}$ diameter dish with a sensitivity of $\mathcal{O}(10^{-19}\,{\rm W})$, which is the strength of the signal from Voyager-1 around 150 AU from Earth detected by the Deep Space Network (DSN), a PBH of mass $M=10^{26}\,{\rm g}$ would have to be at a distance $b\lesssim 9 \, {\rm km}$ from the dish to be detected. 

Instead of relying on an Earth-based radio receiver, one might instead imagine a much larger receiving station. As discussed in Refs.~\cite{Silk:2020bsr,Schneider:2022loy,Silk:2025znp,Zhang:2025eeh}, the dark side of the Moon is a radio-quiet candidate site for future large-scale instruments. In a hypothetical scenario in which the \emph{entire} dark side of the Moon was instrumented to have an effective collecting area $\sim (1700 \,{\rm km})^2$ in the radio band (akin to efforts like the Event Horizon Telescope \cite{EventHorizonTelescope:2019uob}), the maximum detectable impact parameter for an optimally emitting PBH with $M=10^{26} \,{\rm g}$ would still only be $b_{\rm max}\simeq 3\times10^{-3} \,{\rm AU}$. We therefore do not pursue radio-band signals as a viable detection strategy now or in the reasonable future, and do not perform any transit simulation studies.

\section{PBH Explosions}
\label{sec:Explosions}

In the previous section, we focused on detecting photon signals from massive ($M\gtrsim10^{15}\,{\rm g}$), long-lived, quiescent PBHs transiting through the inner Solar System. We found that such asteroid-mass PBHs can transit through the inner Solar System frequently on human time-scales given extended mass function constraints on $f_{\rm PBH}$. In this section, we consider photon signals from a distinct and significantly more rare phenomenon: the rapid, violent explosion of a PBH near the end of its evaporation lifetime. 

For the purposes of this analysis, we define an exploding PBH to have present-day mass $M\lesssim10^{11}\,{\rm g}$, which is hot enough to democratically emit all SM degrees of freedom. PBHs exploding today must have formed with initial mass $M_i= M_*\simeq 5.4\times10^{14}\,{\rm g}$. Note that a PBH of mass $M=5.93\times10^{10}\,{\rm g}$ has a remaining lifetime of $1 \, {\rm day}$, so these explosions are rapid, transient, and extremely high-energy phenomena; in other words, a black hole evaporates gradually, and then suddenly \cite{SunRises}. See Fig.~\ref{fig:HawkExplosion} for secondary photon Hawking emission spectra for PBHs in the explosive mass range $10^{-5}\,{\rm g}\leq M \leq 10^{11}\, {\rm g}$. Emitted photon energies span a range from $E_\gamma \sim \mathcal{O}(10\,{\rm MeV})$ up to the Planck scale $E_\gamma\sim10^{18}\, {\rm GeV}$. 

Given the small-mass tail of the generalized critical collapse PBH number distribution function, $\phi_{\rm GCC} (M_i)$ in Eq.~\eqref{eqn:PhiGCC}, there will generically exist a small subpopulation of PBHs in the present day with masses $M\lesssim10^{11}\,{\rm g}$, leading to a nontrivial number of PBH explosions per cubic parsec per year in the neighborhood of the Solar System. Given the strongest experimental constraint on local PBH burst rates~\cite{LHAASO:2025kyn}, which limits the local PBH burst rate to $\dot{n}\leq 181 \, {\rm pc}^{-3}{\rm yr}^{-1}$ (99\% CL), and direct calculations of PBH explosion rates from realistic number distributions~\cite{Klipfel:2025jql}, explosions relatively near Earth can only be expected to occur on human time-scales in the far outer reaches of the Solar System, with $b \gtrsim 10^3 \,{\rm AU}$. The likelihood of an explosion occurring with $b \ll 10^3 \, {\rm AU}$ over a time-scale of decades or centuries is vanishingly small. 

Assuming rare, nearby explosions are governed by Poisson statistics, one can infer the probability of a PBH explosion occurring within the Solar System during some window of time. For example, assuming an underlying local burst rate of $\dot{n}= 181 \, {\rm pc}^{-3}{\rm yr}^{-1}$, the probability of observing one explosion in a 15-year window within a spherical volume centered on the Sun with radius $b\leq 10^4 \, {\rm AU}$ is 35\%, whereas the probability of observing one event in the same time window in the inner Solar System (within $b\leq 5 \, {\rm AU}$) is $1.6\times10^{-10}$. We therefore only consider photon detection prospects for PBH explosions within the distant reaches of the Solar System, including the Kuiper Belt and Oort Cloud, which extend to $\mathcal{O}(10^3\,{\rm AU})$ and $\mathcal{O}(10^5\,{\rm AU}\sim 1 \, {\rm pc})$ respectively.   

As discussed in Refs.~\cite{Klipfel:2025jql,Boccia:2025hpm,Baker:2025zxm,Baker:2025cff,Anchordoqui:2025xug,Airoldi:2025opo,Airoldi:2025bgr,Ambrosone:2026djo,Mukhopadhyay:2026lmz}, local PBH explosions are candidate sources for ultrahigh-energy cosmic rays, such as the $220 \, {\rm PeV}$ KM3-230213A neutrino event recently reported by the KM3NeT collaboration \cite{aielloObservationUltrahighenergyCosmic2025}. In Ref.~\cite{Klipfel:2025jql}, we analyzed such a scenario and found that the KM3-230213A event was compatible with a rare, local PBH explosion at a distance $b \simeq 1890 \, {\rm AU}$ from Earth. Moreover, we demonstrated that PBH explosions throughout the galactic dark-matter halo, drawn from the same underlying PBH population with $f_{\rm PBH} \simeq 1$, could {\it also} account for the reported IceCube diffuse isotropic neutrino background for $E_\nu \gtrsim {\cal O} (1 \, {\rm PeV})$. Ref.~\cite{Klipfel:2025jql} thus describes a feasible scenario---consistent, within $2 \sigma$, with both the reported IceCube fluxes and LHAASO PBH burst-rate constraints---in which PBHs comprise an ${\cal O}(1)$ fraction of the galactic dark matter, generate the as-yet unexplained PeV-scale neutrino background, and provide a viable transient point source for the KM3-230213A event, which reduces the tension between the IceCube and KM3NeT observations \cite{Li:2025tqf}.

In Section~\ref{sec:BurstsGeneral} we consider possible electromagnetic signatures from such scenarios, in which PBHs explode within the Oort Cloud, with $b \sim {\cal O} (10^3 - 10^5 \, {\rm AU})$. We make a point to emphasize, in particular, the photon signals expected from an explosion of the sort that could also source an
ultrahigh-energy neutrino detection on Earth. As in the previous sections, we consider the standard Hawking-radiation formalism and restrict attention to SM degrees of freedom. We calculate primary and secondary photon emission over a wide range of photon energies from PBH explosions and compute expected signals at Earth for Fermi-LAT, HAWC, and LHAASO: instruments which collectively span ${\rm MeV-PeV}$ photon energy scales. We aim to highlight the real possibility of making a multimessenger detection of a PBH explosion in the outer reaches of the Solar System, an event which has a reasonable likelihood of occurring on human time-scales and is violent enough to generate a measurable signal at such distances.

Then in Section ~\ref{sec:BurstKM3}, we analyze the expected EM counterpart signal to the specific PBH burst scenario which may have sourced the KM3-230213A event \cite{Klipfel:2025jql}. We revisit this event carefully and discuss expected $\gamma$-ray signals for the LHAASO and HAWC observatories. We note that HAWC was offline at the time of the KM3-230213A event~\cite{HAWCtelegram}, and the burst associated with the ultrahigh-energy neutrino occurred outside the field of view of LHAASO~\cite{Airoldi:2025opo}. However, we estimate the possible signal LHAASO could have observed 9 hours \emph{before} the ultrahigh-energy neutrino detection, when the KM3-230213A point source had most recently been within its FOV. At that time, the distant PBH would have been emitting particles including photons at significantly lower energies compared to its final explosion.

\subsection{PBH Burst Duration and Expected Signals}
\label{sec:BurstsGeneral}

A given cosmic ray experiment with energy range $[E_{\rm min}, E_{\rm max}]$ will be sensitive to PBH \emph{bursts} of duration $\tau_{\rm burst} \leq \tau_{\rm burst}^{\rm max}$, which approximately extends from when the PBH reaches a mass $M$ such that $E_{\rm peak}(M)\simeq E_{\rm min}$ until its complete evaporation. The lifetime of a PBH with mass $M\ll M_* \simeq  5.4\times10^{14} \,{\rm g}$ is given by
\begin{equation}
    \tau(M)=\frac{M^3}{3 \mathcal{A}F_{\rm max}},
    \label{eq:tauSmall}
\end{equation}
where ${\cal A}$ and $F_{\rm max}$ are both described near Eq.~(\ref{Mstar}). Combining Eq.~(\ref{EpeakHawk}) and Eq.~(\ref{eq:tauSmall}), the expected maximum burst duration for an instrument is
\begin{equation}
    \begin{split}
    \tau_{\rm burst}^{\rm max}(E_{\rm min}) & = \left(\frac{6.04}{8 \pi G E_{\rm min}} \right)^3\frac{1}{3 \mathcal{A}F_{\rm max}}\\
    & = 3.41 \, {\rm yr} \left(\frac{E_{\rm min}}{100 \, {\rm GeV}} \right)^{-3},
    \end{split}
    \label{eq:TauBurstMax}
\end{equation}
which corresponds to the lifetime of a PBH with remaining mass
\begin{equation}
    M_{\rm burst}^{\rm max}=\frac{6.04}{8 \pi G E_{\rm min}} = 6.39\times10^{11}\,{\rm g}\left(\frac{E_{\rm min}}{100 \, {\rm GeV}} \right)^{-1}.
    \label{eq:MBurstMax}
\end{equation}
Thus the energy range of an instrument determines the PBH mass range it is sensitive to probing, namely $M\leq M_{\rm max}^{\rm burst}$.

Performing burst searches with duration longer than $\tau_{\rm burst}^{\rm max}$ would result only in accumulating additional background counts in the search window, and therefore burst searches often choose $\tau_{\rm burst} \ll \tau_{\rm burst}^{\rm max}$, to minimize background, which is lowest for high energy photons. See Ref.~\cite{LHAASO:2025kyn} for the most recent constraints on the local PBH burst rate with LHAASO, and particularly see Fig.~2 for a summary of burst rate constraints from various cosmic ray observatories~\cite{Linton:2006yu,Alexandreas:1993zx,Glicenstein:2013vha, Archambault:2017asc,Fermi-LAT:2018pfs,Abdo:2014apa,HAWC:2019wla,HESS:2023zzd, LHAASO:2025kyn}. Table~\ref{tab:bursts} lists several instruments sensitive to high-energy photons with their respective energy ranges and maximum burst duration sensitivities. To estimate expected signals for each instrument, we use the burst durations $\tau_{\rm burst}$ for searches reported in the literature (listed in column 6 of Table~\ref{tab:bursts}). Specifically, we analyze the expected signals from a $10\,{\rm s}$ duration burst, which is the largest value searched for by the HAWC Collaboration \cite{HAWC:2019wla}.

\begin{table*}[t!]
\caption{\label{tab:BurstVals} \justifying PBH explosion burst signal parameters. The instrument-dependent photon energy range is $[E_{\rm min}, \, E_{\rm max}]$, with associated maximum burst duration $\tau_{\rm burst}^{\rm max}$ and PBH mass $M_{\rm burst}^{\rm max}$ given by Eqs.~\eqref{eq:TauBurstMax}--\eqref{eq:MBurstMax}.  Note that the maximum burst duration for Fermi-LAT is $\mathcal{O}(400 \, {\rm Gyr})\gg t_0$ due to its sensitivity to lower-energy photons and therefore to PBHs with $M>M_*$; thus, it does not make sense to define the maximum burst duration for this instrument. The largest values of $\tau_{\rm burst}$ used by PBH burst searches in the literature are given in column 6 with associated masses $M_{\rm burst}$ in column 7 and references in column 8. Column 9 reports $N_{\gamma}$, the number of photons with energy $E_{\rm min}\leq E_{\gamma}\leq E_{\rm max}$ emitted in the final $10 \,{\rm s}$ burst weighted by detection efficiency. Column 10 reports the estimated number of \emph{signal} photons measured by each detector according to Eq.~\eqref{eq:Nobs} for a $10 \, {\rm s}$ burst at $b=1890 \,{\rm AU}$.
}
\begin{ruledtabular}
\begin{tabular}{cccccccccc}
  & $E_{\rm min}$ & $E_{\rm max}$ & $\tau_{\rm burst}^{\rm max}$ & $M_{\rm burst}^{\rm max}$ & $\tau_{\rm burst}$ & $M_{\rm burst}$ & Ref. & $N_{\gamma}$ & $N_{\rm sig}$ \\
\hline
Fermi-LAT & $20 \, {\rm MeV}$ & $300 \, {\rm GeV}$ & -- & -- & $3.2 \, {\rm yr}$ & $6.25\times10^{11}\,{\rm g}$ & \cite{Fermi-LAT:2018pfs} & $1.66\times10^{30}$ & $1.7$\\

HAWC & $300 \, {\rm GeV}$ & $10 \, {\rm TeV}$ & $0.13 \, {\rm yr}$& $2.13\times10^{11}\,{\rm g}$& $10 \, {\rm s}$& $2.89\times10^9\,{\rm g}$& \cite{HAWC:2019wla} & $6.38\times10^{28}$ & $6.35\times10^3$ \\

LHAASO-WCDA & $100 \, {\rm GeV}$ & $30 \, {\rm TeV}$ & $3.4 \, {\rm yr}$& $6.39\times10^{11}\,{\rm g}$& $100\,{\rm s}$&$6.23\times10^9\,{\rm g}$ & \cite{LHAASO:2025kyn} &$2.92\times10^{28}$ & $3.47\times10^4$
\label{tab:bursts}
\end{tabular}
\end{ruledtabular}
\end{table*}

\begin{figure}[t]
    \includegraphics[width=0.49\textwidth]{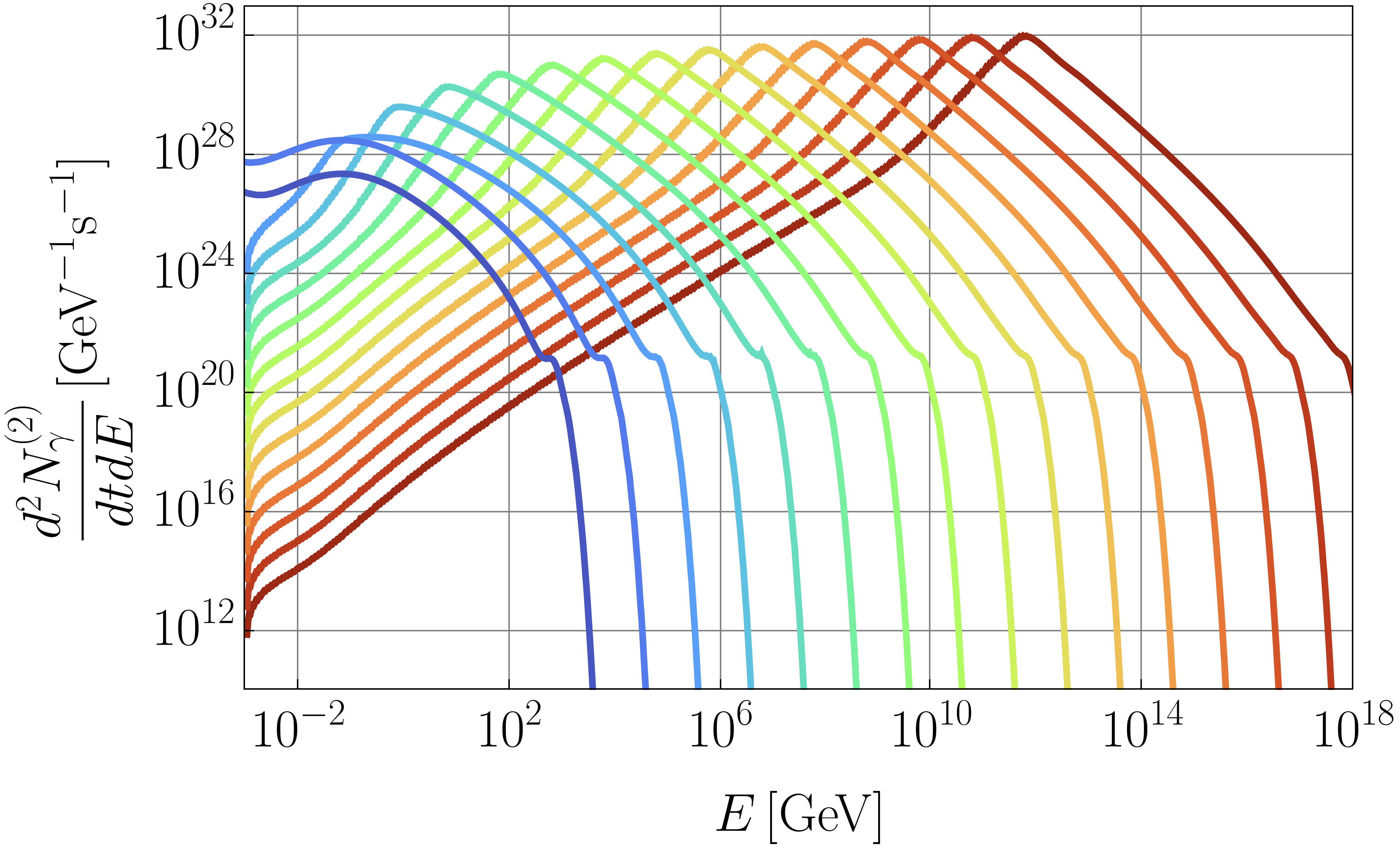} 
    \caption{\justifying Secondary photon emission spectra for PBHs with masses from $10^{-5}\,{\rm g}$ (dark red) to $10^{11}\,{\rm g}$ (dark blue). An exploding PBH will traverse this entire mass range in approximately 5 days, with its mass evolution given by Eq.~\eqref{eq:Moft}. }
    \label{fig:HawkExplosion}
\end{figure}

Figure~\ref{fig:HawkExplosion} shows the secondary photon Hawking emission spectra for PBHs in the explosive mass range $M\leq 10^{11}\,{\rm g}$. As a PBH explodes, its mass decreases as
\begin{equation}
    M(t|M_i)\simeq(M_i^3-3 \mathcal{A}F_{\rm max}t)^{1/3}.
    \label{eq:Moft}
\end{equation}
An exploding PBH of mass $M(t)$ will emit photons with energies $E_\gamma\lesssim E_{\rm peak}= 6.04\, T_H(M(t))$, thus, although the maximum emitted photon energy may be extremely large, the tail of the secondary spectrum admits emission of lower-energy photons at high rates. Hence, we expect Fermi-LAT, HAWC, and LHAASO each to be sensitive to the final $10 \, {\rm s}$ burst, even though they detect photons in different energy ranges. 

Unlike the PBH transits studied in Section~\ref{sec:PhotonSignals}, these final, explosive bursts are rapid events, and are effectively instantaneous on time-scales of the Earth's rotation. As emphasized by Refs.~\cite{Airoldi:2025opo,Airoldi:2025bgr,Mukhopadhyay:2026lmz}, the relevant instrument must be pointing at the correct location in the sky within the short ($\sim 10 \, {\rm s}$) time interval in order to successfully detect a signal from a PBH burst as the energetic particles arrive at Earth. Additionally, note that in this section we are  only interested in high-energy photon detection, so the following analysis should apply to both space-based and ground-based instruments as the Earth's magnetic field can be neglected---unlike in Ref.~\cite{Klipfel:2025bvh}. 

\subsubsection{LHAASO}

The Large High Altitude Air Shower Observatory (LHAASO) consists of two main detectors with distinct energy ranges: the $1.3 \, {\rm km}^2$ array (KM2A) and the $7.8\times10^4\,{\rm m}^2$ water Cherenkov detector array (WCDA) \cite{Ma:2022aau}. Following Ref.~\cite{LHAASO:2025kyn}, we will focus on PBH burst detection with the WCDA, which is expected to be mre sensitive to PBH bursts than KM2A.

\begin{figure}[t]
    \includegraphics[width=0.49\textwidth]{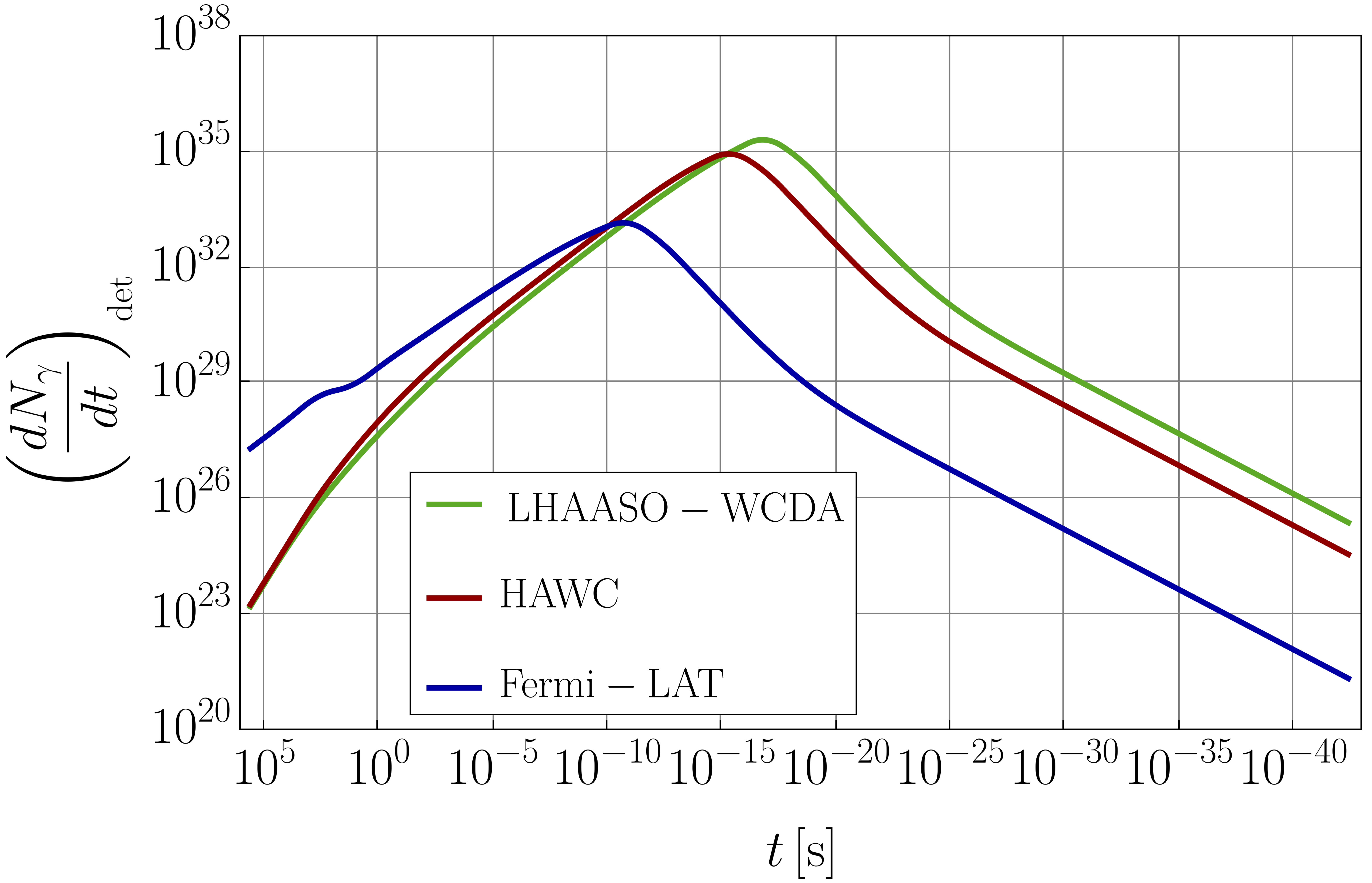} 
    \caption{\justifying Detectable photon emission rate as a function of time throughout a PBH explosion. The PBH is initialized at $t_i=4.13\times10^5\,{\rm s}$ with $M_i=10^{11} \, {\rm g}$ and the explosion completes when the PBH reaches $M=10^{-5}\,{\rm g}$ at $t=4.13\times10^{-43}\,{\rm s}$. Time counts down (from left to right) toward the final explosion. The entire simulated duration is $\tau(10^{11}\,{\rm g})=4.78\,{\rm days}$. }
    \label{fig:BurstEmitRates}
\end{figure}

Using the reported WCDA effective area for different values of zenith angle $\theta$ from Ref.~\cite{Ma:2022aau}, we compute the weighted emission rate of detectable photons in the WCDA energy range $100 \, {\rm GeV}\leq E_{\gamma} \leq 30 \, {\rm TeV}$ via
\begin{equation}
    \left(\frac{dN_\gamma}{dt} \right)_{\rm det} = \int_{E_{\rm min}}^{E_{\rm max}} dE \, \frac{A_{\rm eff}(E, \theta)}{A_{\rm eff}^{\rm max}(\theta)}\frac{d^2N_\gamma^{(2)}}{dEdt}.
    \label{eq:burstdNdtDet}
\end{equation}
See Fig.~\ref{fig:BurstEmitRates} for the detectable emission rate as a function of time. Note that we can express $(dN/dt)_{\rm det}$ as a function of time instead of mass in this regime via Eq.~\eqref{eq:tauSmall}.

We compute $N_{\gamma}$, the efficiency-weighted number of photons with energy $E_{\rm min}\leq E_{\gamma}\leq E_{\rm max}$ emitted in the final $10 \,{\rm s}$ burst for the instrument, via
\begin{equation}
    N_{\gamma}=\int_0^{10 \,{\rm s}}dt \left(\frac{dN_{\gamma}}{dt}\right)_{\rm det}.
    \label{eq:NgammaBurst}
\end{equation}
As in Fig.~\ref{fig:BurstEmitRates}, the explosion completes at $t=0$, so this captures the particle emission during the final $10 \, {\rm s}$. See column 9 of Table~\ref{tab:BurstVals}.

The measured \emph{signal} photon count rate at Earth due to a $10 \, {\rm s}$ burst at some distance $b$ away, for a given instrument with maximum effective area $A_{\rm eff}^{\rm max}$, is thus
\begin{equation}
    N_{\rm sig}(b, \theta) = \frac{N_{\gamma}(\theta)}{4 \pi b^2}A_{\rm eff}^{\rm max}(\theta).
    \label{eq:Nobs}
\end{equation}
See column 9 of Table~\ref{tab:BurstVals} for the values $N_{\rm sig}$ given $\theta\simeq 0$ and $b=1890\,{\rm AU}$  (the proposed distance to a PBH explosion which could have sourced the KM3-230213A event~\cite{Klipfel:2025jql}). 

Ref.~\cite{LHAASO:2025kyn} performed a PBH burst search across all gamma-like events looking for significant excesses above the cosmic ray background via a spatial grid-search, which involves sliding a rectangular grid across the sky map and recording photon counts in each grid over a time window $\Delta t= 10\,{\rm s}, \,20\,{\rm s, \, or}\, 100\,{\rm s}$. Ref.~\cite{LHAASO:2025kyn} used a spatial grid of size $1.2^\circ\times1.2^\circ$, which we will assume to be the smallest region of the sky to which we are capable of resolving a PBH point source. Note that this is a fair assumption given the LHAASO point-spread function (the radius in degrees which encloses 68\% of source photons) for various shower sizes ranges from $0.21^\circ$ to $0.84^{\circ}$~\cite{LHAASO:2021ozi}.

To evaluate the prospects of detecting a PBH explosion with LHAASO, we assume, much like with GRB detection, that the PBH burst has been detected and localized to a point in the sky by another instrument (such as when the KM3 collaboration announced the KM3-230213A event originated from ${\rm RA} = 94.3^\circ$, ${\rm dec.} = -7.8^\circ$ at time ${\rm MJD} = 59988.0533299$~\cite{aielloObservationUltrahighenergyCosmic2025}), and we could then perform a follow-up search with LHAASO data from the time of the observation. Assuming optimistically that a PBH burst candidate has been localized to a $1.2^\circ\times1.2^\circ$ region of the sky, we can compute the signal count threshold for a $5\sigma$ detection above background. Note that instances of poorer localization on the sky would lead to higher backgrounds and thus lower SNR, so this is an upper bound estimate on a detection scenario, and optimistic in the sense of lower background.

\begin{figure*}[t]
    \includegraphics[width=0.49\textwidth]{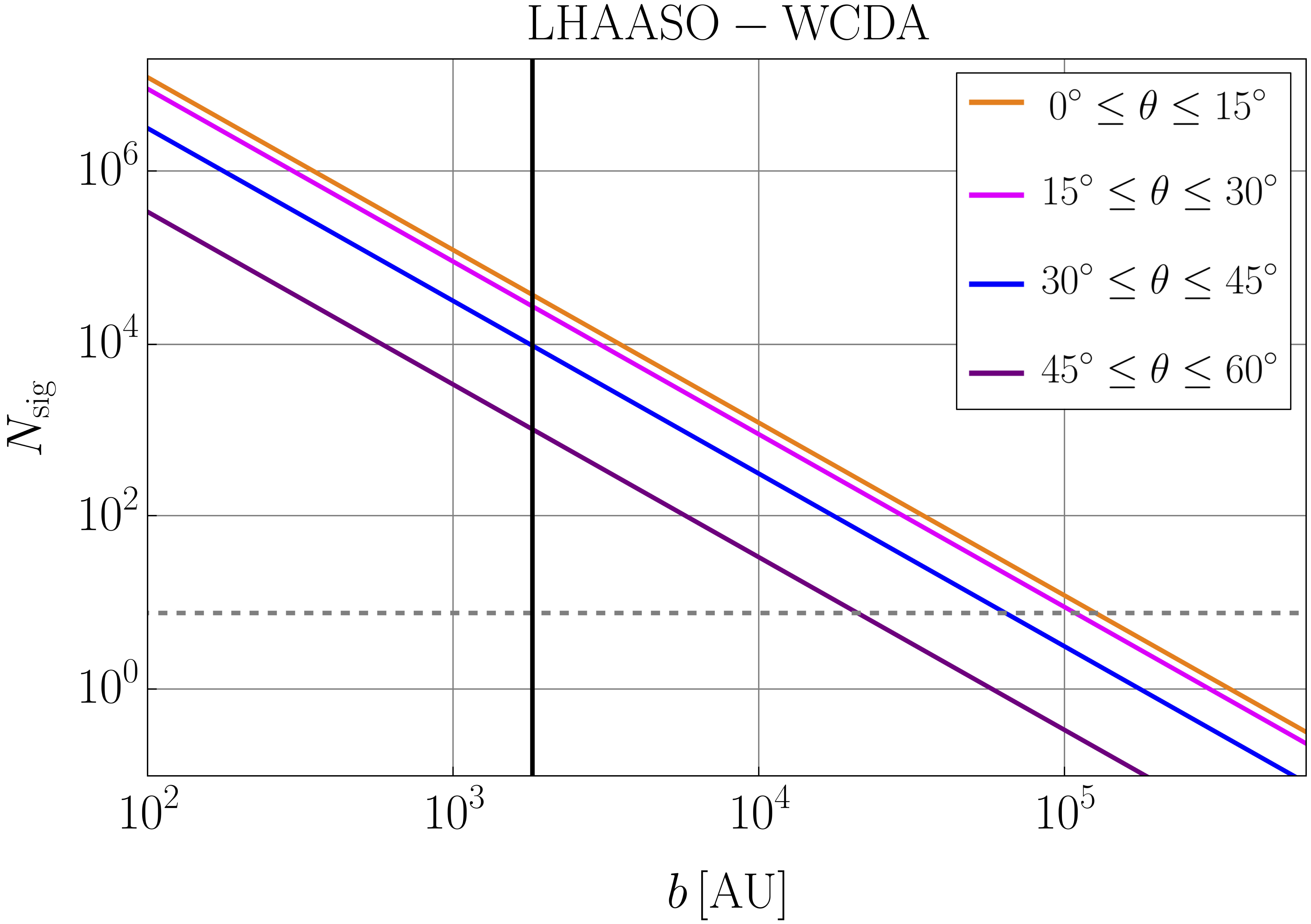} 
    \includegraphics[width=0.49\textwidth]{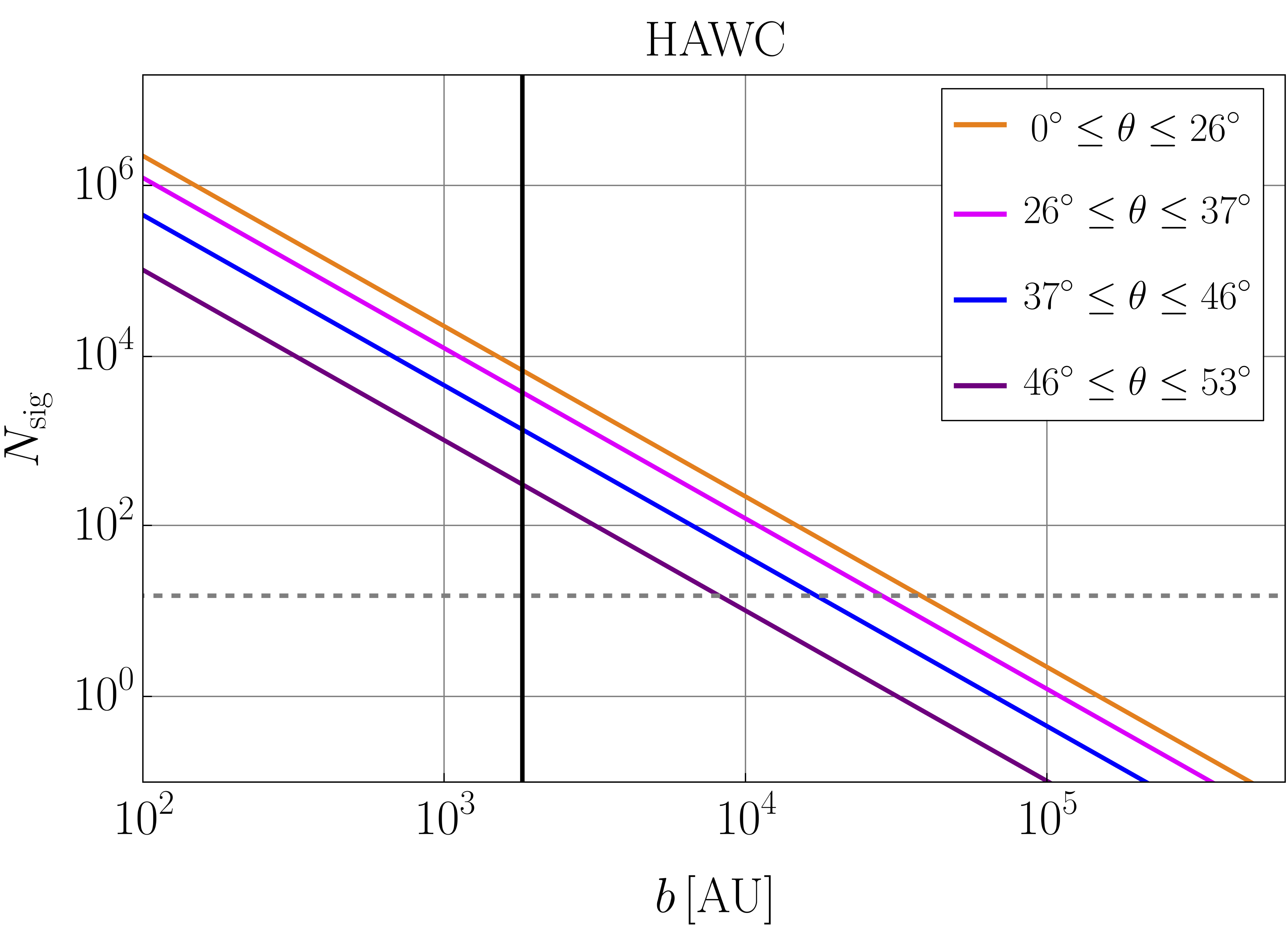}
    \caption{\justifying Measured signal photon counts $N_{\rm sig}(b, \theta)$ for a PBH explosion a distance $b$ from Earth with duration $\tau_{\rm burst}=10\,{\rm s}$ and zenith angle $\theta$, for the LHAASO-WCDA detector ({\it left}) and for HAWC ({\it right}). Both instruments are less sensitive to bursts farther from the zenith $\theta=0$. The vertical black line indicates $b=1890\,{\rm AU}$, the estimated distance for an explosion which could theoretically have sourced the KM3NeT event \cite{Klipfel:2025jql}. The horizontal dashed lines indicate the $5\sigma$ detection threshold for $N_{\rm sig}$ given expected grid backgrounds. Note that for LHAASO-WCDA, the dashed gray line is the $5\sigma$ signal threshold given the FOV-averaged estimated background of $2.5$ counts per $1.2^\circ\times1.2^\circ$ grid \cite{LHAASO:2025kyn}, while for HAWC the dashed grey line is the $5\sigma$ threshold assuming the expected grid background at zenith of 10 counts per $2.1^\circ\times2.1^\circ$ grid \cite{Wood:2016lqc}.}
    \label{fig:LHAASOcounts}
\end{figure*}

To determine the feasibility of detecting a burst with a known position on the sky with signal photon count $N_{\rm sig}(b, \theta)$ using the WCDA, we must compare with the expected background for a $1.2^\circ\times1.2^\circ$ grid. Using the expected background count rates and grid size from Ref.~\cite{LHAASO:2025kyn}, and assuming the WCDA instantaneous FOV covers $1/7$ of the sky, we estimate a photon background of $2.5$ counts per grid in a $10 \, {\rm s}$ window. Note that this estimate assumes backgorund counts are isotropic across the FOV, when in reality background is highest near the zenith.  Assuming  grid background counts are Poisson distributed with mean $\lambda=2.5$, the $5\sigma$ threshold on $N_{\rm sig}$ is thus $5\sqrt{\lambda}\simeq 7.9$ counts.

Figure~\ref{fig:LHAASOcounts} compares the expected signal photon count for a $10\,{\rm s}$ PBH burst, $N_{\rm sig}(b)$, to the $5\sigma$ threshold (dashed line). We find that $N_{\rm sig}(b)$ for a burst near the zenith is above the $5\sigma$ threshold for $b\leq b_{\max}^{\rm WCDA}=0.61 \, {\rm pc}$. This estimate is consistent with Ref.~\cite{LHAASO:2025kyn}, which states that no observable signals from PBH bursts at distances $b\gtrsim 1\, {\rm pc}$ are expected for LHAASO. The vertical black line in Fig.~\ref{fig:LHAASOcounts} indicates $b=1890 \, {\rm AU}$.
We therefore expect a rare PBH explosion at $b\sim10^3\,{\rm AU}$ to be detectable by LHAASO \emph{if the explosion occurs within the detector FOV} at any value of $\theta$.

As a very conservative counterpoint to the above analysis, we can also consider a simple scenario in which the PBH burst with signal photon count $N_{\rm sig}$ occurs in the WCDA FOV, but is \emph{not} assumed to be localized on the sky. The total expected background photon count across the entire WCDA FOV in a $10 \, {\rm s}$ window is $N_{\rm bkgd}=5.09\times10^3$ \cite{LHAASO:2025kyn}. The SNR in this scenario, in which the burst location is unknown, is thus 
\begin{equation}
    {\rm SNR}(b, \theta)=\frac{N_{\rm sig}(b, \theta)}{\sqrt{N_{\rm sig}(b, \theta)+N_{\rm bkgd}}}.
    \label{eq:burstSNR}
\end{equation}
Given $N_{\rm sig}(b=1890 \, {\rm AU}, \theta\simeq0)=3.47\times10^4$ from Table~\ref{tab:BurstVals}, we find ${\rm SNR}(b=1890\,{\rm AU})=174$, 
implying that a burst detection with the WCDA at $b=1890 \, {\rm AU}$ is possible even without prior knowledge of the source location. We find that ${\rm SNR}\geq 1$ for $b\leq 0.2 \, {\rm pc}$ for this conservative detection scenario.

\subsubsection{HAWC}

We repeat the above analysis for the High Altitude Water Cherenkov (HAWC) Observatory, with sensitivity to photons in the range $300 \, {\rm GeV}\lesssim E_\gamma \lesssim 10 \, {\rm TeV}$. Using the reported effective area $A_{\rm eff}(E)$ from Ref.~\cite{HAWC:2011gts}, we first compute the weighted emission rate of detectable photons $(dN/dt)_{\rm det}$ via Eq.~\eqref{eq:burstdNdtDet}. See Fig.~\ref{fig:BurstEmitRates} for a plot of $(dN/dt)_{\rm det}$ as a function of time throughout an explosion. Then we compute the total weighted number of emitted detectable photons $N_\gamma$ by integrating $(dN/dt)_{\rm det}$ over a $10 \, {\rm s}$ burst via Eq.~\eqref{eq:NgammaBurst}. From Eq.~\eqref{eq:Nobs} we then compute the estimated signal photon count rate $N_{\rm sig}$ for a $10 \, {\rm s}$ PBH burst at $b=1890\,{\rm AU}$ along the HAWC line-of-sight. See columns 9 and 10 of Table~\ref{tab:BurstVals}.

As in the previous subsection, we assume the PBH burst is known to be localized to some area on the sky and then determine the $5\sigma$ detection threshold given the expected background count rate for the grid. Ref.~\cite{HAWC:2019wla} reports a PBH burst spatial grid search using a grid size of $2.1^\circ\times2.1^\circ$ and a $50^\circ$ FOV cone. We thus assume the burst is known to be localized to some $2.1^\circ\times2.1^\circ$ region on the sky. Ref.~\cite{Wood:2016lqc} reports an expected background count rate of $\mathcal{O}(1)$ shower per $1\,{\rm s}$ window near zenith for a $2.1^\circ\times2.1^\circ$ grid, implying an expected average background of $\sim10$ counts for a $10\,{\rm s}$ burst. Assuming the background grid counts are Poisson distributed with mean $\lambda=10$, then the $5\sigma$ threshold for detection of a PBH burst with HAWC would be $N_{\rm sig}=5\sqrt{\lambda}=15.8$ counts. We take this as the detection threshold. See the dashed grey line in Fig.~\ref{fig:LHAASOcounts}, which plots the signal $N_{\rm sig}(b)$ measured by HAWC for explosions at different zenith angles $\theta$. We find that, for an explosion near the zenith ($\theta=0$), the maximum detectable impact parameter is $b_{\rm max}^{\rm HAWC}=0.18 \, {\rm pc}$. Ref.~\cite{HAWC:2019wla} reports that HAWC does not expect to detect PBH explosions beyond $\mathcal{O}(0.5\,{\rm pc})$, roughly consistent with our estimate. 

As in the previous subsection, we can also consider a very conservative detection scenario in which the PBH burst location is not known. Assuming a HAWC FOV of $2\,{\rm sr}$ \cite{Springer:2016xzh} and approximately $10$ background counts per $2.1^\circ\times2.1^\circ$ grid, we estimate a total background of $1.49\times10^4$ counts across the whole sky in a $10\,{\rm s}$ window.  Defining the SNR as in Eq.~\eqref{eq:burstSNR}, we find ${\rm SNR}(b=1890\,{\rm AU})=43.6$ and that ${\rm SNR}\geq1$ for $b\leq 0.06 \, {\rm pc}$.

\subsubsection{Fermi-LAT}

We have shown that LHAASO-WCDA and HAWC are viable candidate experiments to detect a rare PBH explosion within the Kuiper Belt or Oort Cloud, if such an explosion occurs within their respective fields of view. However, as noted by Refs.~\cite{Airoldi:2025opo,Airoldi:2025bgr}, the HAWC and LHAASO fields of view do not overlap with each other. We briefly comment on the possibility that the Fermi Large Area Telescope (Fermi-LAT) could coincidentally observe a PBH explosion in the distant reaches of the Solar System along with one of these other two instruments. Such a coincident observation could be possible due to the large $2.4\,{\rm sr}$ Fermi-LAT field of view. The Fermi-LAT instrument is in low-Earth orbit with an orbital inclination of $25^\circ$---which takes its trajectory near the zeniths of both LHAASO (zenith $\sim 30^\circ$ north) and HAWC (zenith $\sim 20^\circ$ north). See Table~\ref{tab:BurstVals} for relevant Fermi-LAT parameters. 

We find that the measured signal count by Fermi-LAT for a $10 \, {\rm s}$ duration, near-zenith PBH burst at $b=1890 \, {\rm AU}$ is $N_{\rm sig}\simeq1.7$ photons, integrated across the entire sensitivity window $20 \, {\rm MeV} \leq E_\gamma \leq 300 \, {\rm GeV}$. We estimate the expected background counts within the Fermi-LAT sensitivity band by integrating the reported $\gamma$-ray background flux $\Phi(E)$ with units ${\rm MeV}^{-1}{\rm cm}^{-2}{\rm s}^{-1}{\rm sr}^{-1}$ weighted by the detector acceptance $a(E)$ with units ${\rm cm}^{2}\,{\rm sr}$:
\begin{equation}
    N_{\rm bkgd}=\tau_{\rm burst}\int_{E_{\rm min}}^{E_{\rm max}}dE \, \Phi(E)\, a(E).
\end{equation}
We take $\Phi$ as the extragalactic $\gamma$-ray background (EGB) flux from Ref.~\cite{Fermi-LAT:2014ryh} (which dominates the diffuse isotropic $\gamma$-ray flux also reported in the same reference) and the Fermi-LAT acceptance from Ref.~\cite{FermiSite}. We find an estimated background count $N_{\rm bkgd}\simeq2.7$ photons and an expected SNR of ${\rm SNR}\simeq0.8<1$. Thus, we do not expect that Fermi-LAT would be able to resolve this $10 \, {\rm s}$ PBH burst of interest at $b=1890 \, {\rm AU}$, nor do we expect that Fermi-LAT could provide a coincident measurement with one of the ground-based observatories for such an event. Fermi-LAT would be able to resolve PBH explosions with $b\ll 10^3 \, {\rm AU}$, but explosions so close to the Earth are extremely low-probability events.

\subsection{Electromagnetic Counterparts to KM3-230213A?}
\label{sec:BurstKM3}

The origins of ultrahigh-energy (UHE) cosmic rays ($E\gtrsim 10^2 \, {\rm PeV} =  10^8\,{\rm GeV}$) remain unknown. The KM3-230213A event, a neutrino with energy $E_\nu\sim \mathcal{O}(10^2 \, {\rm PeV)}$ and no identified astrophysical source \cite{aielloObservationUltrahighenergyCosmic2025}, could possibly have been emitted by an exploding PBH~\cite{Klipfel:2025jql,Boccia:2025hpm,Baker:2025zxm,Baker:2025cff,Anchordoqui:2025xug,Airoldi:2025opo,Airoldi:2025bgr,Ambrosone:2026djo,Mukhopadhyay:2026lmz}. A PBH explosion emits $1\times10^{20}$ neutrinos and $4\times10^{19}$ photons with energies $E_\nu\gtrsim 10^2 \, {\rm PeV}$ \cite{Klipfel:2025jql}, providing a feasible mechanism to generate UHE cosmic rays. Note that UHE protons are also produced, but at rates $\sim 100$ times lower than neutrino emission rates and thus can be neglected here.

Refs.~\cite{Airoldi:2025opo, Airoldi:2025bgr} considered the possibility of a PBH source for KM3-230213A. The authors approached this scenario by computing the necessary distance between the Earth and a PBH explosion such that one neutrino event would be detected by KM3NeT for a $\sim 100 \, {\rm s}$ burst, which roughly amounts to computing the separation distance such that $\mathcal{O}(1)$ UHE neutrino hits every square kilometer on Earth. For such a requirement, they found a required separation distance of $b\simeq 4\times10^{-5}\,{\rm pc}\simeq 8 \, {\rm AU}$. The authors determined that a PBH explosion so close to Earth (roughly at the orbit of Saturn) would generate detectable $\gamma$-ray counterpart signals if the explosion occurred within the FOV of HAWC or LHAASO.
However, Ref.~\cite{Airoldi:2025opo} notes that the point source for the KM3-230213A neutrino identified by Ref.~\cite{aielloObservationUltrahighenergyCosmic2025} lay outside the LHAASO FOV at the time of the KM3-230213A event detection and that HAWC, whose FOV the point source did lie within, was offline at the time of the KM3-230213A detection
and thus unable to detect any coincident signal
\cite{HAWCtelegram}. Nonetheless, the authors showed that an explosion at $b\simeq 4\times10^{-5}\,{\rm pc}$ would have generated a strong, measurable signal in LHAASO 9 hours \emph{prior} to the hypothesized burst, when the PBH had most recently fallen within the LHAASO FOV. Since no such prior signal was observed in LHAASO, the authors determined that such a PBH burst at $b\simeq 4\times10^{-5}\,{\rm pc}$ could not have reasonably been the source of the KM3-230213A event. 

As shown in Sec.~\ref{sec:BurstsGeneral}, a PBH explosion within the inner Solar System would clearly generate an extremely strong signal detectable by HAWC and LHAASO (see Fig.~\ref{fig:LHAASOcounts}). Lack of observation of a coincident $\gamma$-ray signal from the origin point of the KM3-230213A neutrino thus strongly disfavors an explosion at a distance $b\simeq 4\times10^{-5}\,{\rm pc}$ as the source of the KM3-230213A neutrino. 
Moreover, a PBH explosion in the inner Solar System is further disfavored by experimental constraints on local PBH burst rates. As noted above, assuming the LHAASO upper bound on PBH burst rates of $181 \, {\rm pc}^{-3}{\rm yr}^{-1}$, the probability of one explosion within $5 \, {\rm AU}$ of Earth in a $15\,{\rm yr}$ time window is $1.6\times10^{-10}$. Thus, we conclude that a PBH explosion \emph{within the inner Solar System} is not a viable candidate for the KM3-230213A event.

In Ref.~\cite{Klipfel:2025jql}, we addressed a complementary question. Rather than computing the separation distance between the Earth and a PBH explosion necessary for a single neutrino from the burst to uniquely hit the KM3NeT ARCA detector (with reported effective area of $\sim 2\times10^2 \, {\rm m^2}$ at $100 \, {\rm PeV}$ \cite{aielloObservationUltrahighenergyCosmic2025}), we analyzed how far from Earth a PBH explosion would have to be for one ultrahigh-energy cosmic ray (neutrino or photon) with energy $E \gtrsim 60 \,{\rm PeV}$ to \emph{hit some instrumented area on Earth} sensitive to such energies.\footnote{The lower bound of $E=60 \, {\rm PeV}$ we used in Ref.~\cite{Klipfel:2025jql} corresponds to the $2\sigma$ lower bound on the energy deposited in the detector by the KM3-230213A event \cite{aielloObservationUltrahighenergyCosmic2025}.} This better matches the observed scenario of one UHE neutrino detection by KM3NeT and no other simultaneous observations of UHE neutrinos or photons by any other instruments. The analysis in Refs.~\cite{Airoldi:2025opo,Airoldi:2025bgr}, which requires one emitted neutrino to hit KM3NeT, is a \emph{significantly} more restrictive requirement on the separation distance $b$ than requiring only one UHE particle to hit \emph{any} relevant detector.

Our analysis proceeds in three steps. The first step 
was to compute the underlying local PBH explosion rate consistent with the diffuse isotropic neutrino flux at ultrahigh energies \cite{Klipfel:2025jql}.  We
began by considering every known neutrino detection with energy $E_\nu \geq 1 \, {\rm PeV}$, which yielded six events (five at IceCube and one at KM3NeT) over a roughly 15-year time-span, corresponding to the time during which the IceCube observatory has been active. To date, none of these events has been associated with a known astrophysical source and none has any reported associated companion signatures, such as high-energy photons. The diffuse isotropic neutrino flux at $\mathcal{O}(10^2 \, {\rm PeV})$ extrapolated from IceCube data is at $\gtrsim 3 \sigma$ tension with the inferred diffuse isotropic flux from the single KM3NeT event.
Following Ref.~\cite{Li:2025tqf} we thus assumed that the KM3NeT neutrino could be attributed to a transient point source. Given the reported IceCube flux at $\mathcal{O}(10^2 \, {\rm PeV})$,
we then inferred a local PBH explosion rate $\dot{n} \simeq 1.41^{+0.80}_{-0.71} \times 10^3 \, {\rm pc}^{-3} \, {\rm yr}^{-1}$ \cite{Klipfel:2025jql}, which is consistent (within $2\sigma$) with the latest LHAASO bound \cite{LHAASO:2025kyn}. 

For the second step of the analysis, we wanted to determine the likelihood of observing an ultrahigh-energy neutrino or photon ($E \sim {\cal O} (10^2 \, {\rm PeV})$) from a local PBH burst in a 15-year window,
given the underlying average volumetric burst rate $\dot{n}$ \cite{Klipfel:2025jql}. We approached this with a uniform, flat prior for the sky location of such an explosion.
During the time-span of interest, several UHE cosmic-ray observatories (sensitive to photons or neutrinos) have been operational, with a \emph{combined effective area} greater than $3700 \, {\rm km}^2$ \cite{PierreAuger:2015eyc,TelescopeArray:2012uws,IceCube:2008qbc,KM3NeT:2024paj,Ma:2022aau}. Taking as a conservative estimate $A_{\rm eff}^{\rm total} \sim {\cal O} (2500 \, {\rm km}^2)$, we found that a PBH explosion at a distance $b \simeq 1890 \, {\rm AU} = 9.16 \times 10^{-3} \, {\rm pc}$ would generate (on average) one UHE cosmic ray detection at Earth \cite{Klipfel:2025jql}.\footnote{Note that this distance is $\sim 230$ times further than the distance used in the analysis of Refs.~\cite{Airoldi:2025opo,Airoldi:2025bgr}; the difference stems largely from the difference in effective detector areas considered.} 

\begin{figure}[t]
    \includegraphics[width=0.49\textwidth]{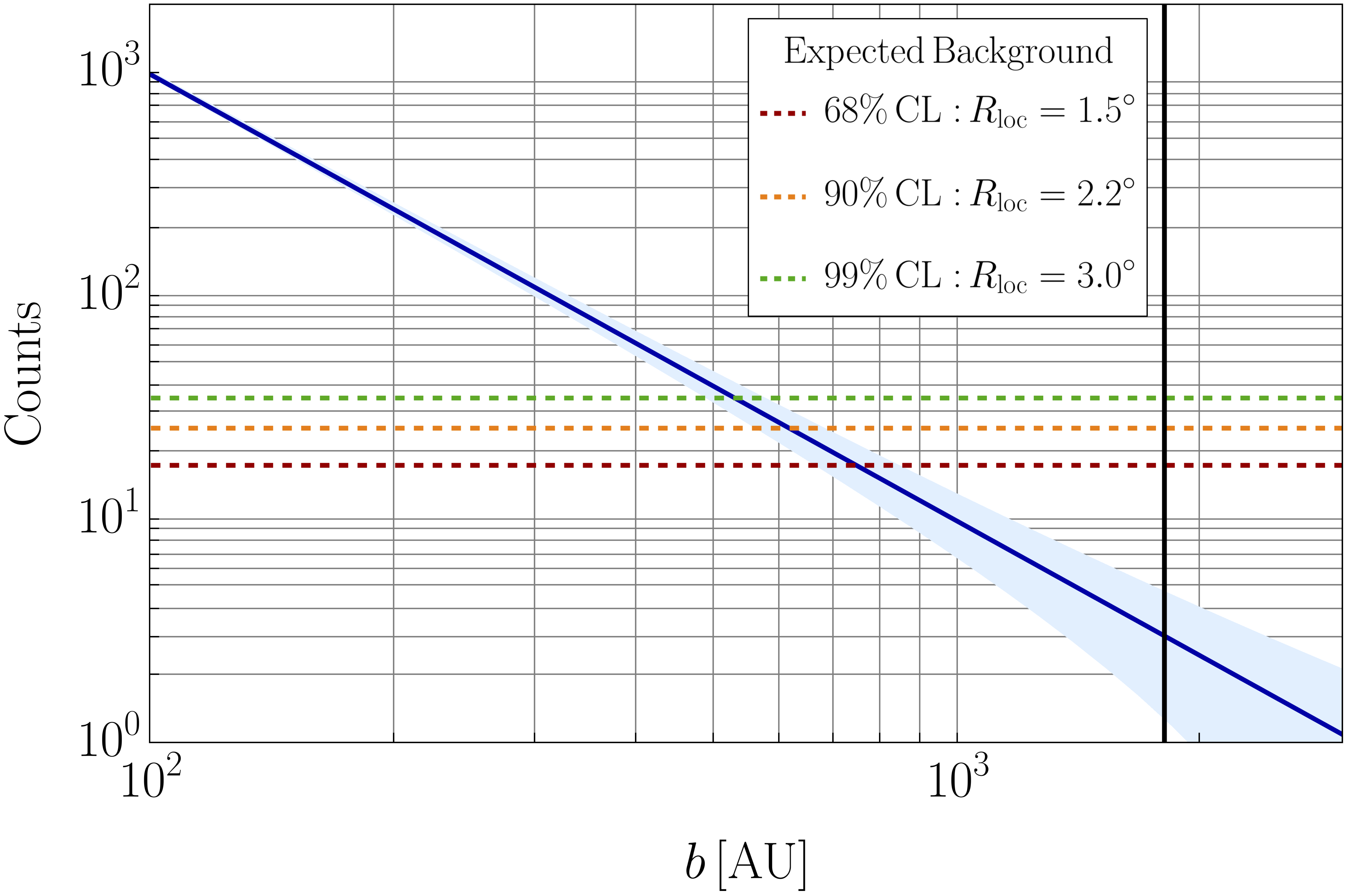} 

    \caption{\justifying Expected photon signal counts (dark blue line) and background counts (dashed lines) for a $10\,{\rm s}$ duration subset of the PBH explosion measured by LHAASO-WCDA 9 hours \emph{prior} to the KM3-230213A event detection---when the KM3-230213A point source was most recently within the WCDA field of view. The $68\%$ confidence interval for the signal count is shown as a light blue band. The hypothesized PBH explosion, at some distance $b$ from Earth, would have been at an angle of $\theta=37.8^\circ$ from zenith for the WCDA \cite{Airoldi:2025opo, aielloObservationUltrahighenergyCosmic2025}. The background counts are estimated for three different levels of point-source localization precision $R_{\rm loc}$ (the radius of area on the sky with some likelihood of containing the point source) reported for the KM3-230213A event from Ref.~\cite{aielloObservationUltrahighenergyCosmic2025}. The solid black vertical line indicates a separation distance of $b=1890\,{\rm AU}$ from Earth, the predicted distance for $\mathcal{O}(1)$ UHE particles from the explosion to hit an instrumented region on Earth \cite{Klipfel:2025jql}. We find that LHAASO would not have been able to detect photon emission from such an explosion 9 hours prior to the final burst (which itself occurred outside the WCDA FOV \cite{Airoldi:2025opo}).}
    \label{fig:WCDA-9hr}
\end{figure}
As discussed in Ref.~\cite{Klipfel:2025jql}, given some underlying PBH explosion rate $\dot{n}$, one may determine the probability for a PBH explosion to occur at a distance $b\leq 1890 \, {\rm AU}$ from Earth within a $\sim 15$-year window. Depending on whether one assumes a burst rate of $\dot{n}\sim\mathcal{O}(200 \, {\rm pc}^{-3}{\rm yr}^{-1})$ \cite{LHAASO:2025kyn} or $\dot{n}\sim\mathcal{O}(2000 \, {\rm pc}^{-3}{\rm yr}^{-1})$ \cite{Klipfel:2025jql}, the probability of one PBH explosion at a distance $b\leq1890 \, {\rm AU}$ from Earth within a 15-year window is $\sim {\cal O}( 1-10\%)$. Thus, given reasonable PBH burst rates consistent with the tightest experimental constraints, there is a nontrivial probability that a PBH drawn from a realistic population within the Milky Way galaxy would explode anomalously close to Earth, such that ${\cal O} (1)$ UHE neutrino or photon would be detected by one of the large detectors that has been operational over that time-span. This scenario could provide a transient point source for the KM3-230213A event if we define the event as just one UHE particle hitting an instrumented area on Earth within a 15-year time-span.

The final step in our analysis of the KM3-230213A event builds on Sec.~\ref{sec:BurstsGeneral} of this work. Given the specific detection event, we
update our prior on the sky location of a possible local PBH burst and assume an explosion occurred $b=1890\,{\rm AU}$ from Earth at the sky location ${\rm RA} = 94.3^\circ$, ${\rm dec.} = -7.8^\circ$ at time ${\rm MJD} = 59988.0533299$~\cite{aielloObservationUltrahighenergyCosmic2025}.
Much as collaborations such as LIGO and IceCube do upon detecting a rare event localized to a particular region of the sky, we then use the updated information to inform searches for possible companion signals---such as observable $\gamma$ rays---from such an event.

There are four main candidate instruments for observation of a $\gamma$-ray signature from this particular event:
LHAASO-WCDA, HAWC, the Pierre Auger Observatory, and the Telescope Array. Ref.~\cite{Airoldi:2025bgr} indicates that the KM3-230213A point source fell within the FOV of the Pierre Auger Observatory at the time of the final burst; however, Ref.~\cite{Tesic:2015gpk} shows that Auger would expect a signal $\sim 10^5$ times weaker than that detected by HAWC. Based on Fig.~\ref{fig:LHAASOcounts}, we therefore do not expect that Auger could have detected a possible $\gamma$-ray counterpart signal. Additionally, Ref.~\cite{Airoldi:2025bgr} shows that the KM3-230213A point source was at too low of a declination to ever have entered the Telescope Array FOV. We thus only focus on the possibility of a photon signal detection with HAWC and LHAASO.

Assuming the burst occurred at distance $b=1890\,{\rm AU}$ with ${\rm RA} = 94.3^\circ$, ${\rm dec.} = -7.8^\circ$ at time ${\rm MJD} = 59988.0533299$~\cite{aielloObservationUltrahighenergyCosmic2025}, the final explosion would have occurred outside the LHAASO-WCDA FOV; and, as previously noted, HAWC was offline at the time. Thus, the final burst would not have been observed by either instrument. However, as pointed out by Ref.~\cite{Airoldi:2025opo}, the hypothesized exploding PBH would have been within the LHAASO-WCDA FOV 9 hours prior to its complete evaporation. We evaluate detection prospects for this signal 9 hours prior to the final burst in Fig.~\ref{fig:WCDA-9hr}. For an explosion at $b=1890\,{\rm AU}$ from Earth, and given the localization precision reported by Ref.~\cite{aielloObservationUltrahighenergyCosmic2025}, we find that the $\gamma$-ray emission from the exploding PBH would \emph{not have been detectable} by LHAASO 9 hours prior to the final burst that generated the KM3-230213A event. Thus there are no observational grounds from photon detection by either HAWC or LHAASO-WCDA to exclude the scenario proposed in Ref.~\cite{Klipfel:2025jql}.

\section{Discussion}

Primordial black holes (PBHs) remain fascinating theoretical objects in their own right, and serve as viable candidates to contribute some or all of the dark matter abundance. Given various constraints, a population of PBHs with typical mass within the asteroid-mass range, $10^{17} \, {\rm g} \lesssim \bar{M} \lesssim 10^{23} \, {\rm g}$, could contribute an ${\cal O} (1)$ fraction of the dark matter density \cite{carrConstraintsPrimordialBlack2021,Carr:2020xqk,greenPrimordialBlackHoles2021,Escriva:2022duf,gortonHowOpenAsteroidmass2024,carrObservationalEvidencePrimordial2024}. Moreover, a population of such PBHs, distributed throughout the Milky Way galaxy in a way that traces the measured dark-matter density, would yield a reasonable transit rate trough the inner Solar System on human time-scales, ranging from once per century to ten times per year \cite{tranCloseEncountersPrimordial2024,Klipfel:2025bvh}. It is therefore of great interest to develop realistic methods to detect PBH transits near Earth.

In this paper, we first focused on electromagnetic Hawking radiation signatures from PBHs transiting through the inner Solar System. We evaluated prospects to detect photons across the UV, X-ray, $\gamma$-ray, and radio bands from local PBH transits with impact parameters $b \lesssim 5 \, {\rm AU}$ from Earth. We simulated time-dependent measured photon signals from PBH transits for existing and proposed instruments (both ground-based and space-based), while taking into account the relative motions of the PBH and the detector, along with the detector efficiencies, orbits, and expected backgrounds. Ideally, measuring photon radiation from a PBH transit would provide a multimessenger signal to complement either measured gravitational perturbations to Solar System objects or the observation of other emitted particles, such as positrons.

PBHs near the upper end of the asteroid-mass range, with $5 \times 10^{20} \, {\rm g} \lesssim M \lesssim 10^{23} \, {\rm g}$, would primarily emit photons within the ultraviolet and infrared, along with nontrivial tails that extend into the radio bands. Although instruments exist that are sensitive to such bands, the relatively low Hawking temperatures $T_H (M)$ of these massive PBHs yield low photon emission rates. Combined with the modest effective areas of relevant instruments for such bands, we find negligible prospects for detecting UV or radio photon radiation from local PBH transits on reasonable time-scales.

Prospects for detecting photon signals from a PBH transit are significantly more favorable in the X-ray and $\gamma$-ray bands. We find that time-series data from the the proposed AMEGO-X satellite, sensitive to photon energies up to $1 \, {\rm GeV}$, would be capable of detecting PBH transits out to impact parameters of $b \lesssim {\cal O} (0.2 \, {\rm AU})$---roughly $10^2$ times greater than the Earth-Moon distance. Notably, however, this is much closer than the typical distances $b \sim {\cal O} (5 \, {\rm AU})$ for which we expect PBH transits to occur on time-scales ranging from once per month to once per decade.

The fact that $\gamma$-rays (and positrons \cite{Klipfel:2025bvh}) from local PBH transits would be observable with existing technology, albeit over relatively long time scales, suggests that is is worthwhile to design novel instruments that admit higher event rates and stronger signals. One might imagine a fleet of detectors, with photon and/or positron sensitivity optimized in the MeV band, distributed across novel orbits---such as Lagrange points or heliocentric orbits---thus increasing the 
PBH encounter rate at distances of $b\leq b_{\rm max}$, while avoiding near-Earth effects such as the geomagnetic field and reducing background from Solar System objects. We leave this intriguing possibility for further research.

An additional point of further research lies in reconstructing PBH properties and trajectories from simulated (and eventually, possibly real) time-dependent photon data from an instrument like AMEGO-X. As we discussed in Ref.~\cite{Klipfel:2025bvh}, matched filter analysis of time-series photon count data using a bank of templates parameterized by $M, v, b$ could allow us to estimate the PBH transit parameters. Correlated transit signatures such as gravitational perturbations \cite{tranCloseEncountersPrimordial2024} may be required to avoid degeneracies in such parameter estimation. Furthermore, if a transit is detected via photons over a sufficiently wide frequency band with high enough statistics, we could imagine performing a fit to the black hole emission spectrum to estimate $M$ and test the thermal nature of the Hawking emission spectrum. We leave this exciting prospect for future research.

In addition to PBH transits through the inner Solar System, we also considered detection prospects for photon signals from PBH explosions in the far reaches of the Solar System---out to $10^3-10^5 \, {\rm AU}$ from Earth. Given realistic extended mass distributions and a PBH dark matter fraction $f_{\rm PBH}\sim\mathcal{O}(1)$, PBH explosions could only occur on human time-scales within a volume of space exponentially larger than the inner Solar System. For example, the probability of one explosion within the Oort Cloud with impact parameter $b\leq10^4\,{\rm AU}$ in a 15 year window is $\sim35\%$.

Unlike typical Hawking radiation from relatively quiescent asteroid-mass PBHs, the particles emitted in the final burst of a PBH explosion can reach ultrahigh energies, up to $E \sim 10^2 \, {\rm PeV} = 10^8 \, {\rm GeV}$ and beyond. Given the typically large distances between Earth and a PBH explosion, relatively few such ultrahigh-energy particles would be expected to reach instrumented areas on Earth.

Multiple large-area cosmic ray observatories have been operational with combined observing times of $\sim 15$ years, including HAWC and LHAASO. We find that both HAWC and LHAASO could detect highly energetic photons from a PBH explosion $b \leq 10^4 \, {\rm AU}$ from Earth, given their effective areas and associated background counts---{\it if} the point-like PBH burst happened to occur within the field of view of either instrument. Given such capability, we analyzed the expected photon signatures associated with a specific event, namely, a PBH explosion at a distance $b \simeq 1890 \, {\rm AU}$ that could have sourced the ultrahigh-energy KM3-230213A neutrino event ($E_\nu \simeq 220 \, {\rm PeV}$) reported by the KM3NeT collaboration \cite{aielloObservationUltrahighenergyCosmic2025, Klipfel:2025jql}. Given the reported sky localization of the KM3-230213A event \cite{aielloObservationUltrahighenergyCosmic2025}, Refs.~\cite{Airoldi:2025opo,Airoldi:2025bgr} found that the final burst of a PBH explosion at that same location would have occurred outside the LHAASO FOV and that no signal could have been reported by HAWC since it was offline at the time. They report that the earliest time that the PBH would have fallen in the LHAASO FOV prior to the final explosion would have been about 9 hours before the KM3-230213A event detection. We find that, if the KM3-230213A event originated from a PBH explosion at an estimated distance of $b \simeq 1890 \, {\rm AU}$ from the Earth, then the photon signal 9 hours prior to the KM3-230213A event detection would have been too weak to be detected by the LHAASO-WCDA.

Thus, if the intriguing KM3-230213A neutrino did arise from a PBH explosion within the Oort Cloud, then no companion photon signal would be expected, given the alignment and operational status of relevant detectors at the time of the event. On the other hand, existing instruments, including both HAWC and LHAASO-WCDA, {\it do} offer realistic detection possibilities for ultrahigh-energy photons from PBH explosions within the Oort Cloud; furthermore, we could expect detectable explosions to occur in the outer reaches of the Solar System every 1-2 decades if PBHs within the asteroid-mass window do constitute a significant fraction of the galactic dark matter. 

In summary, we find that there are viable multimessenger detection strategies for both PBH transits and explosions within our Solar System for mass distributions peaked in the asteroid-mass window with $f_{\rm PBH}\simeq1$. The measurement of Hawking-radiated photon signals would be a valuable component of multimessenger local PBH detection strategies and should be seriously considered when constructing future cosmic-ray observatories and analyzing data from existing instruments.

\section*{Acknowledgements}

We gratefully acknowledge helpful discussions with Michael Baker, Shyam Balaji, Peter Fisher, Joaquim Iguaz Juan, Yuber F.~Perez-Gonzalez, Tracy Slatyer, Aidan Symons, and Andrea Thamm. Portions of this research were conducted in MIT's Center for Theoretical Physics --- A Leinweber Institute and supported by the Office of High Energy Physics within the Office of Science of the U.S.~Department of Energy under grant Contract Number DE-SC0012567. This material is based upon work supported by the National Science Foundation Graduate Research Fellowship under Grant No.~2141064. We also gratefully acknowledge support from the Amar G.~Bose Research Grant Program at MIT.


%

\end{document}